\documentclass[aps,pra,twocolumn,a4paper,showpacs]{revtex4}
\usepackage{amsmath}
\usepackage{amssymb}
\usepackage{times,txfonts}
\usepackage{bbold}

\begin{document}

\title{A number-conserving approach to a minimal self-consistent treatment of condensate and non-condensate dynamics in a degenerate Bose gas}
 
\author{S. A. Gardiner}
\affiliation{Department of Physics, Durham University,
Rochester Building, South Road, Durham DH1 3LE, United Kingdom}

\author{S. A. Morgan}
\affiliation{Department of Physics and Astronomy, University College London, 
Gower Street, London WC1E 6BT, United Kingdom}
\altaffiliation[Present address: ]{Lehman Brothers, 25 Bank Street, London E14 5LE, United Kingdom}

\date{\today}

\begin{abstract}
We describe a number conserving approach to the dynamics of Bose-Einstein condensed dliute atomic gases.  This builds upon the works of Gardiner [C. W. Gardiner, 
Phys.\ Rev.\ A \textbf{56}, 1414 (1997)], and Castin and Dum [Y. Castin and R. Dum, 
Phys.\ Rev.\ A \textbf{57}, 3008 (1998)].  We consider what is effectively an expansion in powers of the ratio of non-condensate to condensate particle numbers, rather than inverse powers of the total number of particles.  This requires the number of condensate particles to be a majority, but not necessarily almost equal to the total number of particles in the system.  We argue that a second-order treatment of the relevant dynamical equations of motion is the minimum order necessary to provide consistent coupled
condensate and non-condensate number dynamics for a finite total number of particles, and show that such a second-order treatment is provided by a suitably generalized Gross-Pitaevskii equation, coupled to the Castin-Dum number-conserving formulation of the Bogoliubov-de Gennes equations.  The necessary equations of motion can be generated from an approximate third-order Hamiltonian, which effectively reduces to second order in the steady state.  Such a treatment as described here is suitable for dynamics occurring at finite temperature, where there is a significant non-condensate fraction from the outset, or dynamics leading to dynamical instabilities, where depletion of the condensate can also lead to a significant non-condensate fraction, even if the non-condensate fraction is initially negligible.  
\end{abstract}


\maketitle

\section{Introduction}
\label{Sec:Introduction}

By definition, a dilute atomic gas that has undergone Bose-Einstein condensation \cite{Bose1924,Einstein1924,Einstein1925} has a large number of component particles occupying the same mode \cite{Einstein1924,Einstein1925,Mandl1988,Abrikosov1975,Fetter2003,Pethick2002,Pitaevskii2003}.  Effects associated with such a macroscopic occupation were first observed in superfluid helium and in superconducting metals \cite{Tilley1990}.  The importance of interactions in such comparatively dense condensed-matter systems means that the condensate fraction, although important, is substantially less than the non-condensate fraction.  In systems composed of laser and magnetically cooled and trapped dilute atomic gases \cite{Anderson1995,Davis1995,Bradley1997} the situation is often very different;  the atomic gas can be sufficiently cold and dilute for the condensate fraction to be a large proportion of the total number of atoms.  It is for this reason that the Gross-Pitaevskii equation \cite{Ginzburg1958,Pitaevskii1961,Gross1961,Gross1963}, originally conceived to develop a qualitative understanding of processes in superfluid helium, has achieved the status of a quantitatively useful description of degenerate dilute gases of bosonic atoms.

The Gross-Pitaevskii equation is essentially a classical field approximation to an underlying quantum field.  Notwithstanding its broad utility, there are many situations where a more accurate description is required.  Superfluid to Mott-insulator phase-transitions in optical lattices \cite{Greiner2001}, and dimer formation via controlled manipulation of magnetic fields (in order to exploit Feshbach resonances) \cite{Donley2002} are topical examples of such processes.  Effectively the strength of the inter-atomic interactions becomes significant to the extent that higher-order atom-atom correlations must be more carefully accounted for, and for which the standard Gross-Pitaevskii equation is inadequate \cite{Jaksch1998,Kokkelmans2002,Kohler2003,Duine2003,HutsonReview,KohlerReview}.

Even apart from such extreme situations, if the non-condensate fraction becomes significant, a description going beyond the Gross-Pitaevskii equation must be called upon.  Two important situations where this may occur are: dynamics occurring at a (significant) finite temperature \cite{Jin1997}, of interest due to the unique possibility offered by dilute Bose-Einstein condensate experiments for quantitative tests of thermal field theories; and dynamics leading to dynamical  instabilities in, and hence depletion of an initially low temperature condensate \cite{Castin1997,Gardiner2000,SGardiner2002}, such as may well occur in experiments \cite{Ryu2006,Behinaein2006,Duffy2004} studying chaotic and quantum chaotic dynamics in Bose-Einstein condensates \cite{Xie2005,Zhang2004,Liu2006,Salmond2002,Hai2002,Martin2006}.   The desire to provide a relatively simple, consistent description of condensate and non-condensate dynamics motivates the work presented here, and a form of the approach we present was a key part in work carried out  \cite{Morgan2003,Morgan2004,Morgan2005}, to good agreement with experiment \cite{Jin1997}, in order to describe excitations at finite temperature of a dilute Bose-condensed gas.

The first recourse when wishing to go beyond the Gross-Pitaevskii equation \cite{Morgan2003,Morgan2004,Morgan2005,Bogoliubov1947,deGennes1966,Fetter1972,Gardiner1997,Castin1998,Girardeau1959,Girardeau1998,Beliaev1958a,Beliaev1958b,Hugenholtz1959,Popov1965a,Popov1965b,Shi1998,Fedichev1998,Mohling1960a,Mohling1960b,Morgan2000} is frequently the Bogoliubov, or Bogoliubov-de Gennes equations \cite{Bogoliubov1947,deGennes1966,Fetter1972}, or their number-conserving variants \cite{Gardiner1997,Castin1998,Girardeau1959,Girardeau1998}.  Particularly motivated by the desire to explain the properties of Bose-condensed gases at finite temperature, a number of extensions have been proposed.  These include generalizations \cite{Giorgini2000,Minguzzi1997,Guilleumas1999,Rusch1999,Rusch2000,Bene1998,Reidl2000}
of linear response theory \cite{Ruprecht1996,Edwards1996}, stochastic interpretations of the Gross-Pitaevskii equation \cite{Davis2001a,Davis2001b,Stoof2001,Duine2001,Gardiner2002}, Hartree-Fock-Bogoliubov approaches \cite{Griffin1996,Hutchinson1997,Dodd1998,Reidl1999,Proukakis1998a,Hutchinson1998,Hutchinson2000,Zhang2006,Zhang2007}, a variety of kinetic theories \cite{Bijlsma1999,AlKhawaja2000,Zaremba1998,Nikuni1999,Zaremba1999,Jackson2001,Jackson2002a,Jackson2002b,Walser1999,Walser2000,Wachter2002,Proukakis2001}, and a cumulant-based formalism \cite{Kohler2003,Kohler2002,KohlerLetter,Goral2004}.

The description presented here is within a number-conserving formalism, and builds on the works of Gardiner \cite{Gardiner1997}, and Castin and Dum \cite{Castin1998}, which are essentially equivalent to each other.  Symmetry-breaking formulations, which automatically violate particle number conservation, have met with considerable success in describing the observed properties of Bose-Einstein condensed dilute atomic gases.  Here symmetry-breaking is defined as the breaking of the $U(1)$ global phase symmetry whose Noether charge is the total particle number.  Technically, however, symmetry-breaking formulations require a coherent superposition of different numbers of particles.  One could argue that the actual particle number is only known statistically in any real experiments, and should be considered an ensemble average from multiple realizations of the same experiment. Even given this, it is difficult to see how shot-to-shot number-coherences could be built up.  It is therefore important to understand any differences which might appear between number-conserving and symmetry-breaking formulations.  

A significant difference that does occur, in the number-conserving formulation presented in this paper, is the presence of nonlocal terms in the equations of motion for the condensate and non-condensate fractions.  These arise from defining the two fractions in such a way that they must be mutually orthogonal. This orthogonality is not in general fulfilled if one defines the condensate fraction as being the expectation value of a field operator, as occurs in symmetry-breaking formulations.  The nonlocal terms can have a significant effect; correct inclusion of these terms has been observed to be crucial in obtaining good agreement between application of the number-conserving formulation presented in this paper \cite{Morgan2003,Morgan2004,Morgan2005}, and the experimental study of collective excitations in a finite-temperature Bose-Einstein condensate \cite{Jin1997}.

We consider what is effectively an expansion in powers of the ratio of non-condensate to condensate particle numbers, rather than inverse powers of the total number of particles \cite{Gardiner1997,Castin1998}.  This means that there should be significantly more condensate than non-condensate particles, but that the condensate fraction is not necessarily assumed to be nearly encompassing the entire many-body system. We argue that a second-order treatment (in the dynamical equations of motion) is the minimum order necessary to provide consistent condensate and non-condensate number dynamics, with exchange of particles between the fractions, for a finite total number of particles.  We show that such a second-order treatment is provided by a suitably generalized Gross-Pitaevskii equation, coupled to the Castin-Dum number-conserving formulation of the Bogoliubov-de Gennes equations (these are modified only by the presence of projectors necessary to maintain orthogonality between the condensate and non-condensate components) \cite{Castin1998}.  The necessary equations of motion can be generated from an approximate third-order Hamiltonian, which effectively reduces to second order in the steady state.  

This paper is organized as follows: Section \ref{Sec:System} formally describes the many-Boson system under consideration, and determines a suitable fluctuation operator on which to base the expansion; Section \ref{Sec:Construction} constructs an appropriate cubic approximate Hamiltonian used to generate the desired equations of motion, and justifies the approximations made;  Section \ref{Sec:Equations} elucidates the equations of motion detailing both condensate and non-condensate dynamics, systematically to zeroth, first, and second order in the fluctuation operators; Section \ref{Sec:Equilibrium} discusses some considerations when the system is assumed to be in an equilibrium state; and Section \ref{Sec:Conclusions} consists of the conclusions. There then follow two technical appendices elaborating on points made in the main text, which are included for completeness.

\section{System properties}
\label{Sec:System}

\subsection{Model Hamiltonian}
\label{SubSec:Model}

The starting point of the theory is the binary interaction Hamiltonian for a system of bosons,
\begin{equation}
\begin{split}
\hat{H}(t) =& \int d\mathbf{r} 
\hat{\Psi}^{\dagger}(\mathbf{r})
H_{\mbox{\scriptsize sp}}(\mathbf{r},t)
\hat{\Psi}(\mathbf{r})
\\&
+ \frac{U_0}{2} \int d\mathbf{r} 
\hat{\Psi}^{\dagger}(\mathbf{r}) 
\hat{\Psi}^{\dagger}(\mathbf{r})
\hat{\Psi}(\mathbf{r})
\hat{\Psi}(\mathbf{r}).
\end{split}
\label{Eq:FieldHamiltonian}
\end{equation}
The field operators obey the standard bosonic commutation relations
$[\hat{\Psi}(\mathbf{r}),
\hat{\Psi}^{\dagger}(\mathbf{r'})] = \delta(\mathbf{r}-\mathbf{r'})$,
and
\begin{equation}
H_{\mbox{\scriptsize sp}}(\mathbf{r}) = -\frac{\hbar^2}{2m} \nabla^2 + V(\mathbf{r})
\label{Eq:ParticleHamiltonian}
\end{equation} 
(where $m$ is the atomic mass) is the single-particle Hamiltonian, containing the kinetic energy and any external potential energy terms. 

For simplicity we have assumed the binary interaction to be characterized by an energy-independent contact potential, 
$V_{\mbox{\scriptsize bin}}(\mathbf{r}-\mathbf{r'}) = U_0 
\delta(\mathbf{r}-\mathbf{r'})$, where 
$U_0 = 4\pi\hbar^2a_{s}/m$ and $a_{s}$ is the $s$-wave scattering length. This is a standard approximation for three-dimensional, cold dilute Bose gases. As is well known, however, it leads to ultra-violet divergences, which are removed by renormalizing various quantities appearing in the subsequent development of the theory. This procedure is well understood and has been discussed by a number of authors (see, for example, Refs.\ \cite{Hutchinson2000,Morgan2000,Rusch1999,Stoof1993,Proukakis1998,Kohler2002,Bijlsma1997}). It can be rigorously justified, and we give a brief outline of the relevant arguments in Appendix \ref{App:Renormalize}.

\subsection{Condensate and fluctuation terms}
\label{SubSec:Condensate}

\subsubsection{Single-body density matrix}
\label{SubSubSec:Single-Body}
In the same manner as Castin and Dum \cite{Castin1998}, we follow Penrose and Onsager \cite{Penrose1956} in defining the condensate wavefunction.  This is in terms of the single-body density matrix 
\begin{equation}
\rho (\mathbf{r}, \mathbf{r'},t) = \langle \hat{\Psi}^{\dagger}(\mathbf{r'})
\hat{\Psi} (\mathbf{r})
\rangle.
\label{Eq:DensityMatrix}
\end{equation}
The single-body density matrix is Hermitian, i.e., $\rho (\mathbf{r}, \mathbf{r'},t)^{*}=\rho (\mathbf{r'}, \mathbf{r},t)$,  and can be decomposed into a complete set of eigenfunctions with real eigenvalues.  As $\rho (\mathbf{r}, \mathbf{r'},t)$ may be time-dependent, these are the instantaneous eigenfunctions and eigenvalues, defined by the diagonal representation of the single-body density matrix at a specific time $t$.

We assume that there is one distinct eigenfunction $\phi (\mathbf{r},t)$, defined with unit norm, which has a corresponding eigenvalue $N_{c}(t)$ significantly larger than all the other eigenvalues.  Thus  
\begin{equation}
\int d \mathbf{r'}
\rho (\mathbf{r}, \mathbf{r'},t)\phi(\mathbf{r'},t) = 
N_{c}(t)\phi(\mathbf{r},t)
\label{Eq:CondensateEigenvalue}
\end{equation}
 (time arguments are used to indicate a possible explicit time dependence).  We are free to partition the field operator into condensate and non-condensate components:
\begin{equation}
\hat{\Psi} (\mathbf{r}) = \hat{a}_{c}(t)\phi(\mathbf{r},t) + \delta\hat{\Psi}(\mathbf{r},t),
\label{Eq:FieldPartition}
\end{equation}
where $\hat{a}_{c}(t)$ annihilates a particle in mode $\phi(\mathbf{r},t)$, and $\delta\hat{\Psi}(\mathbf{r},t)$ is that part of the field operator $\hat{\Psi} (\mathbf{r})$ orthogonal to $\phi(\mathbf{r},t)$.  We refer to $\phi (\mathbf{r},t)$ as the condensate mode.

\subsubsection{Comparison with symmetry-breaking}
\label{SubSubSec:Symmetry}
The conventional symmetry-breaking methodology to partition the field operator into condensate and non-condensate components assumes a finite expectation value $\Psi(\mathbf{r},t)=\langle \hat{\Psi}(\mathbf{r})\rangle$ for the field operator, and  additionally considers a fluctuation term $\delta\hat{\Psi}_{\mbox{\scriptsize sb}}(\mathbf{r},t)$ around this mean value \cite{Abrikosov1975,Fetter2003,Pethick2002,Pitaevskii2003}.  Hence, one states:
\begin{equation}
\hat{\Psi}(\mathbf{r}) = \Psi(\mathbf{r},t) + \delta\hat{\Psi}_{\mbox{\scriptsize sb}}(\mathbf{r},t).
\end{equation}
By definition, a fluctuation around a mean or expectation value must itself have a mean or expectation value equal to zero, i.e., $\langle  \delta\hat{\Psi}_{\mbox{\scriptsize sb}}(\mathbf{r},t)  \rangle = 0$.  It also follows that these fluctuation operators have exactly bosonic commutation relations, i.e.,
\begin{equation}
[\delta\hat{\Psi}_{\mbox{\scriptsize sb}}(\mathbf{r},t),\delta\hat{\Psi}_{\mbox{\scriptsize sb}}^{\dagger}(\mathbf{r'},t)] =
[\hat{\Psi}(\mathbf{r}),\hat{\Psi}^{\dagger}(\mathbf{r'})] 
= \delta(\mathbf{r}-\mathbf{r'}).
\end{equation}

Comparing the partitioned terms arising from a symmetry-breaking formulation with those arising from consideration of the single-body density matrix, we get
\begin{gather}
\Psi(\mathbf{r},t) = \langle  \hat{a}_{c}(t)\rangle\phi(\mathbf{r},t) + \langle \delta \hat{\Psi}(\mathbf{r},t) \rangle,
\label{Eq:Comparison1}
\\
\begin{split}
\delta\hat{\Psi}_{\mbox{\scriptsize sb}}(\mathbf{r},t)
=&
[\hat{a}_{c}(t)-\langle  \hat{a}_{c}(t)\rangle]\phi(\mathbf{r},t) \\
& +
[\delta \hat{\Psi}(\mathbf{r},t)-\langle \delta \hat{\Psi}(\mathbf{r},t) \rangle].
\end{split}
\label{Eq:Comparison2}
\end{gather}
To enable a direct comparison with symmetry-breaking, we have not assumed particle-number conservation, and hence neither $\langle  \hat{a}_{c}(t)\rangle$ nor $\langle \delta \hat{\Psi}(\mathbf{r},t) \rangle$ are automatically assumed $=0$.  It is apparent that exactly which parts of the many-body system are designated as condensate and non-condensate, depends on the formulation employed.  One consequence is that, unlike $\phi(\mathbf{r},t)$ and $\delta\hat{\Psi}(\mathbf{r},t)$, the terms $\Psi(\mathbf{r},t)$ and $\delta\hat{\Psi}_{\mbox{\scriptsize sb}}(\mathbf{r},t)$ arising from a symmetry-breaking formulation are not, in general, spatially orthogonal.  As can be seen from Eq.\ (\ref{Eq:Comparison1}) and Eq. \ (\ref{Eq:Comparison2}), in general they both have contributions from both $\phi(\mathbf{r},t)$ and $\delta \hat{\Psi}(\mathbf{r},t)$. 

From the Penrose-Onsager \cite{Penrose1956} formulation for partitioning the field operator used in this paper, it follows that those parts of the many-body system designated as being condensate and non-condensate must always be orthogonal.  It is this requirement that leads to higher-order (i.e., beyond Gross-Pitaevskii) equations of motion necessarily including nonlocal terms, not arising in symmetry-breaking formulations.  It should be emphasized that this difference does not in itself imply that one formulation is more correct than the other, although partitioning the field operator into manifestly orthogonal components can be a considerable convenience.

\subsubsection{Commutation relations}
\label{SubSubSec:Commutation}
Formally, the condensate mode creation operator $\hat{a}_{c}^{\dagger}(t)$ and the non-condensate field operator $\delta\hat{\Psi}(\mathbf{r},t)$ can be defined with respect to the bosonic field operator $\hat{\Psi}(\mathbf{r})$:
\begin{gather}
\hat{a}_{c}^{\dagger} (t)= \int d\mathbf{r} \hat{\Psi}^{\dagger}(\mathbf{r})
\phi(\mathbf{r},t),
\label{Eq:CondensateCreate}\\
\delta\hat{\Psi}(\mathbf{r},t)=\int d \mathbf{r'} Q(\mathbf{r},\mathbf{r'},t)\hat{\Psi}(\mathbf{r'}), 
\label{Eq:NonCondensateField}
\end{gather}
where the projector $Q(\mathbf{r},\mathbf{r'},t)$ is defined to be
\begin{equation}
Q(\mathbf{r},\mathbf{r'},t) =\delta(\mathbf{r}-\mathbf{r'})-\phi(\mathbf{r},t)\phi^{*}(\mathbf{r'},t).
\label{Eq:Projector}
\end{equation}
It follows that the only non-zero commutation relations involving $\hat{a}_{c}(t)$, $\delta\hat{\Psi}(\mathbf{r},t)$, and their Hermitian conjugates are: 
\begin{gather}
[\hat{a}_{c}(t),\hat{a}_{c}^{\dagger}(t)] = 1, \\
[\delta\hat{\Psi}(\mathbf{r},t),\delta\hat{\Psi}^{\dagger}(\mathbf{r'},t)] = Q(\mathbf{r},\mathbf{r'},t).
\label{Eq: Commutator}
\end{gather}

\subsection{Fluctuation expansion}
\label{SubSec:FluctuationExpansion}

\subsubsection{Overview}
\label{SubSubSec:Overview}

It is a common procedure to consider a mean value of some observable quantity, and to account for the observed variance in the values obtained from multiple runs of some real or hypothetical measurement procedure with a fluctuation term \cite{Gardiner2004}.  The observable quantity is then described in terms of its mean or expectation value, plus a fluctuation term with (by definition) zero expectation value.  If the observable is a simple one-dimensional quantity, such as the vertical position of a classical point-mass, then the statistics of the measured values can be represented by a simple one-dimensional distribution.  

If the fluctuations are vanishingly small, then the distribution of a continuous variable collapses to a $\delta$-function, but otherwise the distribution will have a width, which one usually describes in terms of the second-order cumulant (the variance).  The scale of the fluctuations is therefore set by the square root of the variance (the standard deviation). Conversely, finite fluctuations imply a finite width, and hence a finite variance.  In the case of a Gaussian distribution, all higher-order moments and cumulants can be described in terms of the distribution's mean and variance.

The symmetry-breaking methodology used to partition the field operator, as described in Section \ref{SubSubSec:Symmetry}, is an example of describing an operator in terms of its expectation value plus a fluctuation term.  Analogues to the variance are the second-order cumulants \cite{Kohler2002,Gardiner2004}:
\begin{gather}
\langle
\hat{\Psi}^{\dagger}(\mathbf{r})
\hat{\Psi}(\mathbf{r'})
\rangle
-\Psi^{*}(\mathbf{r},t) 
\Psi(\mathbf{r'},t) 
 = 
\langle
\delta\hat{\Psi}_{\mbox{\scriptsize sb}}^{\dagger}(\mathbf{r},t)
\delta\hat{\Psi}_{\mbox{\scriptsize sb}}(\mathbf{r'},t)
\rangle,
\\
\langle
\hat{\Psi}(\mathbf{r})
\hat{\Psi}(\mathbf{r'})
\rangle
-\Psi(\mathbf{r},t) 
\Psi(\mathbf{r'},t) 
 = 
\langle
\delta\hat{\Psi}_{\mbox{\scriptsize sb}}(\mathbf{r},t)
\delta\hat{\Psi}_{\mbox{\scriptsize sb}}(\mathbf{r'},t)
\rangle,
\end{gather}
together with their Hermitian conjugates.  As $\langle\int d\mathbf{r} \delta\hat{\Psi}_{\mbox{\scriptsize sb}}^{\dagger}(\mathbf{r},t)\delta\hat{\Psi}_{\mbox{\scriptsize sb}}(\mathbf{r},t)\rangle$ is the number of non-condensate particles, we generally consider $\delta\hat{\Psi}_{\mbox{\scriptsize sb}}(\mathbf{r},t)$ to scale as the square root of the number of non-condensate particles.  Note, however, that even if all second- (and higher-) order \textit{normally\/} ordered cumulants equal zero, there still remain purely quantal fluctuation terms, due to the noncommutativity of the field operators.

If the physically relevant matrix elements of the fluctuation terms can be considered to be in some sense small (for example with respect to the square root of the number of condensate particles, as considered in this paper), then one can attempt an expansion of the Hamiltonian [Eq.\ (\ref{Eq:FieldHamiltonian})] and equations of motion in terms of products of the fluctuation terms.  Terms of beyond a specified order can then be discarded as negligible, up to the order of interest \cite{Abrikosov1975,Fetter2003,Pethick2002,Pitaevskii2003,deGennes1966}. 

In this paper, we wish to carry out a comparable program, but within a number-conserving formalism \cite{Gardiner1997,Castin1998,Morgan2000}.  As the expectation values of all particle creation and annihilation operators are (trivially) equal to zero, we cannot define a ``small'' fluctuation operator by taking a field operator and subtracting its (zero) expectation value.  An appropriate fluctuation operator that does not trivially have zero expectation value is required.  The validity of such expansions is in general reliant upon (at least in the homogeneous limit)  $(Na_{s}^{3})^{1/2}\ll 1$ if $T=0$, and $(k_{b}T/N_{c}U_{0})(N_{c}a_{s}^{3})^{1/2} \ll 1$ if $(k_{b}T/N_{c}U_{0})\gg 1$, where $N$ is the total particle number, $T$ is the temperature and $k_{B}$ is Boltzmann's constant \cite{Morgan2000}.

\subsubsection{Number-conserving fluctuation operators}
\label{SubSubSec:Number-Conserving}
Substituting Eq.\ (\ref{Eq:FieldPartition}) into Eq.\ (\ref{Eq:CondensateEigenvalue}) reveals that $N_{c}(t) = \langle \hat{N}_{c}(t)\rangle$, where $\hat{N}_{c}(t) \equiv \hat{a}_{c}^{\dagger}(t)\hat{a}_{c}(t) $. The eigenvalue $N_{c}(t)$ is thus the mean number of particles in the condensate mode.  Furthermore,
\begin{equation}
\langle \hat{a}_{c}^{\dagger} (t) \delta\hat{\Psi}(\mathbf{r},t) \rangle = 0,
\label{Eq:Expectation}
\end{equation}
i.e., there are no simple coherences between the condensate and non-condensate components.

The product $\hat{a}_{c}^{\dagger} (t) \delta\hat{\Psi}(\mathbf{r},t)$ has much to recommend it as a fluctuation operator suited to a number-conserving formalism.  Unlike $\delta\hat{\Psi}(\mathbf{r},t)$, its mean value is not \textit{trivially\/} equal to zero in a number-conserving approach.  The desired number-conserving fluctuation operator should be of the same magnitude as $\delta\hat{\Psi}(\mathbf{r},t)$, however.  

Castin and Dum \cite{Castin1998}, and Gardiner \cite{Gardiner1997} were thus motivated to define 
\begin{equation}
\hat{\Lambda}(\mathbf{r},t) = \frac{1}{\sqrt{\hat{N}}}\hat{a}_{c}^{\dagger}(t)\delta\hat{\Psi}(\mathbf{r},t),
\label{Eq:Lambda}
\end{equation}
where $\hat{N}=\int d\mathbf{r} \hat{\Psi}^{\dagger}(\mathbf{r})\hat{\Psi}(\mathbf{r})$ is the total particle number operator.  
Introducing the notation $N_{t}(t)=N-N_{c}(t)$ for the non-condensate fraction, and noting that $\hat{a}_{c}(t)$ should scale as $\sqrt{N_{c}(t)}$ and $\delta\hat{\Psi}(\mathbf{r},t)$ as $\sqrt{N_{t}(t)}$, it then follows that $\hat{\Lambda}(\mathbf{r},t)$ scales as
\begin{equation}
\sqrt{\frac{N_{c}(t)N_{t}(t)}{N}} = 
\sqrt{N_{t}(t)\left[1-\frac{N_{t}(t)}{N}\right]} \approx
\sqrt{N_{t}(t)} 
\end{equation}
for $N\gg N_{t}(t)$. Hence, under the assumption that $N_{c}(t) \approx N$, this operator scales satisfactorily.

When considering an assembly of exactly $N$ atoms, Eq.\ (\ref{Eq:Expectation}) directly implies that $\langle \hat{\Lambda}(\mathbf{r},t)\rangle \equiv 0$, and so $\hat{\Lambda}(\mathbf{r},t)$ can be considered a simple fluctuation operator.  To linear order in $\hat{\Lambda}(\mathbf{r},t)$ \cite{Castin1998}, 
\begin{equation}
[\hat{\Lambda}(\mathbf{r},t),\hat{\Lambda}^{\dagger}(\mathbf{r'},t)] \approx
[\delta\hat{\Psi}(\mathbf{r},t),\delta\hat{\Psi}^{\dagger}(\mathbf{r'},t)] = Q(\mathbf{r},\mathbf{r'},t),
\label{Eq:LambdaCommutator}
\end{equation}
and
\begin{equation}
\int d\mathbf{r} \hat{\Lambda}^{\dagger}(\mathbf{r},t)\hat{\Lambda}(\mathbf{r},t)
\approx
\int d\mathbf{r} \delta\hat{\Psi}^{\dagger}(\mathbf{r},t)\delta\hat{\Psi}(\mathbf{r},t)
= \hat{N} - \hat{N}_{c}(t).
\label{Eq:NonCondensateNumber}
\end{equation}
Again, when considering an assembly of exactly $N$ atoms, there can be no fluctuations in the total number operator, and so $\hat{N} - \hat{N}_{c}(t)$ can be identified with $N - \hat{N}_{c}(t)$.

We wish to avoid making the assumption that $N_{c}(t) \approx N$, i.e., that almost all bosons are in the condensate mode, and so consider a scaling proportionate to the number of \textit{condensate\/} atoms rather than the total number of atoms \cite{Morgan2000,Morgan2004,Girardeau1959,Girardeau1998}.  

A possible alternative to $\hat{\Lambda}(\mathbf{r},t)$ [Eq.\ (\ref{Eq:Lambda})] is
\begin{equation}
\hat{\Lambda}_{c}(\mathbf{r},t) = \frac{1}{\sqrt{\hat{N}_{c}(t)}}\hat{a}_{c}^{\dagger}(t)\delta\hat{\Psi}(\mathbf{r},t),
\label{Eq:LambdaExact}
\end{equation}
from which the exact identities
\begin{equation}
[\hat{\Lambda}_{c}(\mathbf{r},t),\hat{\Lambda}_{c}^{\dagger}(\mathbf{r'},t)] =
[\delta\hat{\Psi}(\mathbf{r},t),\delta\hat{\Psi}^{\dagger}(\mathbf{r'},t)] = Q(\mathbf{r},\mathbf{r'},t),
\label{Eq:LambdaCommutatorExact}
\end{equation}
and
\begin{equation}
\int d\mathbf{r} \hat{\Lambda}_{c}^{\dagger}(\mathbf{r},t)\hat{\Lambda}_{c}(\mathbf{r},t)
=
\int d\mathbf{r} \delta\hat{\Psi}^{\dagger}(\mathbf{r},t)\delta\hat{\Psi}(\mathbf{r},t)
= \hat{N} - \hat{N}_{c}(t).
\label{Eq:NonCondensateNumberExact}
\end{equation}
follow.  This strong correspondence between normal \textit{pairs\/} of $\hat{\Lambda}_{c}(\mathbf{r},t)$ and $\delta\hat{\Psi}(\mathbf{r},t)$ operators appears very attractive.  However, the expectation value
$\langle \hat{\Lambda}_{c}(\mathbf{r},t) \rangle$
is \textit{not\/} guaranteed to be identically equal to zero.  
Consequently, the operator $\hat{\Lambda}_{c}(\mathbf{r},t)$ cannot be treated as a simple operator-valued fluctuation.  This complicates the development of a consistent expansion in terms of products of $\hat{\Lambda}_{c}(\mathbf{r},t)$  for the determination of improved equations of motion.  

In particular, the methodology introduced by Castin and Dum \cite{Castin1998}, and employed in slightly modified form in Section \ref{Sec:Equations} of this paper, relies explicitly on 
$
\langle d \hat{\Lambda}(\mathbf{r},t)/dt \rangle 
\equiv
d
\langle \hat{\Lambda}(\mathbf{r},t)\rangle/dt
=0
$
for the development of the  equations of motion.  As $\langle \hat{\Lambda}_{c}(\mathbf{r},t)\rangle$ is not exactly equal to zero, an equivalent expression is not guaranteed to hold for $\hat{\Lambda}_{c}(\mathbf{r},t)$.  One could still exploit the idea that $\langle \hat{\Lambda}_{c}(\mathbf{r},t)\rangle$ should be \textit{approximately\/} equal to zero, but this would in principle require a careful accounting of higher-order corrections to such an approximation, for example involving Taylor expanding $[\hat{N}_{c}(t)]^{-1/2}$ in powers of $[\hat{N}_{c}(t)-N_{c}(t)]/N_{c}(t)$.

Equation (\ref{Eq:Expectation}) tells us that $\hat{a}_{c}^{\dagger}(t)\delta\hat{\Psi}(\mathbf{r},t)$ and any \textit{scalar\/} multiple thereof has expectation value exactly equal to zero.  In order to avoid such complications as described above, we thus choose to carry out an expansion in terms of
\begin{equation}
\tilde{\Lambda}(\mathbf{r},t) = \frac{1}{\sqrt{N_{c}(t)}}\hat{a}_{c}^{\dagger}(t)\delta\hat{\Psi}(\mathbf{r},t).
\label{Eq:LambdaTilde}
\end{equation}
The normal $\tilde{\Lambda}(\mathbf{r},t)$ pair is related to the normal $\delta\hat{\Psi}(\mathbf{r},t)$ pair via
\begin{equation}
\tilde{\Lambda}^{\dagger}(\mathbf{r'},t) \tilde{\Lambda}(\mathbf{r},t) =
\frac{\hat{N}_{c}(t)+1}{N_{c}(t)}\delta\hat{\Psi}^{\dagger}(\mathbf{r'},t)\delta\hat{\Psi}(\mathbf{r},t),
\label{Eq:LambdaTildePair}
\end{equation}
and the exact commutation relation is given by
\begin{equation}
\begin{split}
[ \tilde{\Lambda}(\mathbf{r},t),\tilde{\Lambda}^{\dagger}(\mathbf{r'},t)] = &
\frac{\hat{N}_{c}(t)}{N_{c}(t)}Q(\mathbf{r},\mathbf{r'},t) \\&-
\frac{1}{N_{c}(t)}\delta\hat{\Psi}^{\dagger}(\mathbf{r'},t)\delta\hat{\Psi}(\mathbf{r},t).
\end{split}
\label{Eq:LambdaTildeCommutator}
\end{equation}

\subsubsection{Comparison of the candidate fluctuation operators}
\label{SubSubSec:Comparison}

The operators $\tilde{\Lambda}(\mathbf{r},t)$ and $\hat{\Lambda}_{c}(\mathbf{r},t)$, in comparison to $\hat{\Lambda}(\mathbf{r},t)$, scale more straightforwardly as $\sqrt{N_{t}(t)}$.  In the expansion used in this paper, the terms that appear in the Hamiltonian and equations of motion have the form of products of $\tilde{\Lambda}(\mathbf{r},t)/\sqrt{N_{c}(t)}$, and/or its Hermitian conjugate [there is an overall factor of $N_{c}(t)$ in the Hamiltonian, and an overall factor of $\sqrt{N_{c}(t)}$ in the equations of motion for $\tilde{\Lambda}(\mathbf{r},t)$ and $\tilde{\Lambda}^{\dagger}(\mathbf{r},t)$].  The fluctuations must thus be small compared to the square root of the number of condensate particles.  In other words, the small parameter such a fluctuation expansion is $\sqrt{N_{t}(t)/N_{c}(t)}$ \cite{Castin1998}. At the very least, greater than $N/2$ particles must therefore be in the condensate mode for the truncation of higher-order terms to be justifiable.  If $U_{0}N$ is kept constant, then, at zero temperature, $N_{t}$ tends to a constant value as $N\rightarrow\infty$ (the quantum depletion). Hence,
\begin{equation}
\sqrt{\frac{N_{t}(t)}{N_{c}(t)}} \rightarrow
\sqrt{\frac{N_{t}(t)}{N}} \propto
\frac{1}{\sqrt{N}},
\label{Eq:SmallNInf}
\end{equation}
and one can consider the small parameter in an expansion in terms of $\hat{\Lambda}(\mathbf{r},t)$, $\hat{\Lambda}^{\dagger}(\mathbf{r},t)$ to be $1/\sqrt{N}$ \cite{Castin1998}.

In Sections \ref{Sec:Construction} and \ref{Sec:Equations} we will include terms of order $[1/N_{c}(t)]^{k/2}$ whenever we include terms of order $[N_{t}(t)/N_{c}(t)]^{k/2}$ ($k$ an integer).  We are obliged to do this if we wish to include the limiting case described by Eq.\  (\ref{Eq:SmallNInf}) in the formalism, although, as the non-condensate fraction increases, the contribution of such terms relative to terms scaling as $[N_{t}(t)/N_{c}(t)]^{k/2}$ decreases.

In Section \ref{SubSec:SecondOrderEquilibrium} we will see that a direct consequence of Eq.\ (\ref{Eq:LambdaTildeCommutator}) being only \textit{approximately \/} equal to $[\delta\hat{\Psi}(\mathbf{r},t),\delta\hat{\Psi}^{\dagger}(\mathbf{r'},t)] $ is that a quasiparticle  formulation produces quasiparticle creation and annihilation operators that are only approximately bosonic.  This is equally the case for the commutator $[\hat{\Lambda}(\mathbf{r},t),\hat{\Lambda}^{\dagger}(\mathbf{r'},t)]$ [Eq.\ (\ref{Eq:LambdaCommutator})], but not for  $[\hat{\Lambda}_{c}(\mathbf{r},t),\hat{\Lambda}_{c}^{\dagger}(\mathbf{r'},t)] =
[\delta\hat{\Psi}(\mathbf{r},t),\delta\hat{\Psi}^{\dagger}(\mathbf{r'},t)]$ [Eq.\ (\ref{Eq:LambdaCommutatorExact})].  A problem that therefore arises, not present in comparable symmetry-breaking formulations, is that the desire to have a well-defined fluctuation operator does not appear to be fully compatible with the desire for this fluctuation operator to induce bosonic quasiparticle commutation relations.

One could take the view that, as non-zero corrections to $\langle \hat{\Lambda}_{c}(\mathbf{r},t) \rangle$ would only appear at a higher order than is being considered in this paper, one could equivalently consider an expansion in terms of $\hat{\Lambda}_{c}(\mathbf{r},t)$ up to the order of current interest \cite{Morgan2004}.  This effectively erases any difference between $\hat{\Lambda}_{c}(\mathbf{r},t)$ and $\tilde{\Lambda}(\mathbf{r},t)$, however, and it is more straightforward, especially when determining truncations of the many-body Hamiltonian necessary to generate the equations of motion, to consider an expansion in terms of $\tilde{\Lambda}(\mathbf{r},t)$ from the outset.  In particular, doing this avoids the complication of having to approximate square-root condensate number operators to an appropriate order in the fluctuation operators.  This appears to cause particular complications in deriving a correct approximate third-order Hamiltonian compatible with the equations of motion calculated to the corresponding order (although not the approach presented in this paper, the equations of motion can be deduced directly, without first determining an approximate Hamiltonian from which they should be derived, as in Ref.\ \cite{Castin1998}).  This does leave somewhat open the question of what the best approach is if one wishes to extend the theory to include higher-order terms \cite{Walser1999,Walser2000,Wachter2002,Proukakis2001}.

We now summarize the relative merits of $\hat{\Lambda}(\mathbf{r},t)$, $\hat{\Lambda}_{c}(\mathbf{r},t)$, and $\tilde{\Lambda}(\mathbf{r},t)$.  We wish to have a non-trivial fluctuation operator that scales like $\delta\hat{\Psi}(\mathbf{r},t)$, i.e., as $\sqrt{N_{t}(t)}$.  Both $\hat{\Lambda}_{c}(\mathbf{r},t)$, and $\tilde{\Lambda}(\mathbf{r},t)$ have this property, whereas this is only true for $\hat{\Lambda}(\mathbf{r},t)$ in the limit $N_{t}(t)/N\rightarrow 0$.  Of $\hat{\Lambda}_{c}(\mathbf{r},t)$ and $\tilde{\Lambda}(\mathbf{r},t)$,  $\tilde{\Lambda}(\mathbf{r},t)$ is a simple fluctuation operator (expectation value exactly equal to zero) but corresponds to not-exactly-bosonic quasiparticle operators, whereas $\hat{\Lambda}_{c}(\mathbf{r},t)$ corresponds to exactly bosonic quasiparticle operators, but does not have expectation value exactly equal to zero and is therefore not a simple fluctuation operator.  In this paper we have made the decision that, for the development of the equations of motion, it is simpler to consider an expansion in terms of $\tilde{\Lambda}(\mathbf{r},t)$.  In particular, this avoids the difficulty of having to approximate square-root condensate-number operators in a consistent way.

\subsubsection{Fluctuation statistics}
\label{SubSubSec:Fluctuation}
In a similar way to how the first manifestation of a fluctuation about a real mean value is in the variance of its corresponding distribution, for a finite number of particles, the presence of fluctuation operators effectively implies non-zero values for such pair expectation values as $\langle \tilde{\Lambda}^{\dagger}(\mathbf{r'},t) \tilde{\Lambda}(\mathbf{r},t)  \rangle$.  This is, of course, with the important caveat that, as $\tilde{\Lambda}(\mathbf{r},t)$ and $\tilde{\Lambda}^{\dagger}(\mathbf{r'},t)$ do not commute, there will always be quantum fluctuations even if $\langle \tilde{\Lambda}^{\dagger}(\mathbf{r'},t) \tilde{\Lambda}(\mathbf{r},t)  \rangle=0$.  With an interacting gas, however, $\langle \tilde{\Lambda}^{\dagger}(\mathbf{r'},t) \tilde{\Lambda}(\mathbf{r},t)  \rangle$ always has a finite value. 

We thus insist \textit{a priori\/} that equations of motion should be taken to quadratic order in products of the fluctuation operators $\tilde{\Lambda}(\mathbf{r},t)$ and $\tilde{\Lambda}^{\dagger}(\mathbf{r},t)$.  
This paper lays out a quite general framework for the development of higher-order theories  involving both the condensate and non-condensate fractions.  However, we also have a very specific goal, which is to determine a minimally complicated set of equations for the self-consistent treatment of condensate and non-condensate dynamics.  Bearing this in mind, a straightforward simplification is  to enforce that all possible expectation values of products of the fluctuation operators $\tilde{\Lambda}(\mathbf{r},t)$ and $\tilde{\Lambda}^{\dagger}(\mathbf{r},t)$ are either zero (for odd products of fluctuation operators), or expressible in products of pair expectation values.  This is essentially a Gaussian approximation, i.e., one assumes that all cumulants, or connected correlation functions, of order greater than two can be considered negligible \cite{Kohler2002}.  One can in principle extend this approximation to also account for the existence of higher-order cumulants (and hence higher-order correlations).  The quantitative validity of such an approximation will depend on the specific dynamics under consideration.

This in turn implies that the the many-body Hamiltonian [Eq.\ (\ref{Eq:FieldHamiltonian})] should be approximated to cubic order in the fluctuation operators [Eq.\ (\ref{Eq:LambdaTilde})].  This is the minimum order necessary to produce equations of motion to quadratic order, and it is not our intention  in this paper to account for any terms of any higher order.

\section{Construction of a third-order Hamiltonian}
\label{Sec:Construction}

\subsection{Transformation of the full Hamiltonian}
\label{SubSec:Transformation}
Until now, every term with an explicit time-dependence has been shown with a $t$ argument.  From now on we neglect this, but it should be remembered that $\phi(\mathbf{r})$, $\tilde{\Lambda}(\mathbf{r})$, $N_{c}$, $\hat{N}_{c}$, and $Q(\mathbf{r},\mathbf{r'})$, all are in general explicitly time-dependent.

One can readily transform the many-body Hamiltonian of Eq.\ (\ref{Eq:FieldHamiltonian}), by everywhere expanding the field operators according to Eq.\ (\ref{Eq:FieldPartition}), and then collecting terms to produce products of $\tilde{\Lambda}(\mathbf{r})$.  Defining $\tilde{U}=U_{0}N_{c}$, the result of carrying this out is: 
\begin{equation}
\begin{split}
\hat{H} =& N_{c}
\frac{\hat{N}_{c}}{N_{c}}
\int d\mathbf{r} 
\left[
\phi^{*}(\mathbf{r})H_{\mbox{\scriptsize sp}}(\mathbf{r})\phi(\mathbf{r})
+
\frac{(\hat{N}_{c}-1)}{N_{c}}
 \frac{\tilde{U}}{2}|\phi(\mathbf{r})|^{4}
 \right]
\\&+
\sqrt{N_{c}}
\int d\mathbf{r} 
\left[
\phi^{*}(\mathbf{r})H_{\mbox{\scriptsize sp}}(\mathbf{r})\tilde{\Lambda}(\mathbf{r})+
\mbox{H.c.}
\right]
\\&+
\sqrt{N_{c}}\tilde{U}\int d\mathbf{r}
\left[
\phi^{*}(\mathbf{r})|\phi(\mathbf{r})|^{2}\frac{\hat{N}_{c}-1}{N_{c}}\tilde{\Lambda}(\mathbf{r}) 
+\mbox{H.c.}
\right]
\\&+
\int d\mathbf{r} 
\tilde{\Lambda}^{\dagger}(\mathbf{r})
\left[
\frac{N_{c}}{\hat{N}_{c}}H_{\mbox{\scriptsize sp}}(\mathbf{r})
+\frac{\hat{N}_{c}-1}{\hat{N}_{c}}2\tilde{U} |\phi(\mathbf{r})|^{2}
\right]
\tilde{\Lambda}(\mathbf{r})
\\&+
\frac{\tilde{U}}{2}\int d\mathbf{r} 
\left[
\phi^{*}(\mathbf{r})^{2}\tilde{\Lambda}(\mathbf{r})^{2} + \mbox{H.c.} \right]
\\&+
\frac{\tilde{U}}{\sqrt{N_{c}}}
\int d\mathbf{r} 
\left[
\phi^{*}(\mathbf{r})\tilde{\Lambda}^{\dagger}(\mathbf{r})
\frac{N_{c}}{\hat{N}_{c}}\tilde{\Lambda}(\mathbf{r})^{2}
+\mbox{H.c.}\right]
\\&+
\frac{\tilde{U}}{N_{c}}
\int d\mathbf{r} \tilde{\Lambda}^{\dagger}(\mathbf{r})^{2}
\frac{N_{c}^{2}}{\hat{N}_{c}(\hat{N}_{c}-1)}\tilde{\Lambda}(\mathbf{r})^{2}, 
\end{split}
\label{Eq:LambdaTildeHamiltonian}
\end{equation}
where the terms are arranged in ascending order of products of the fluctuation operators $\tilde{\Lambda}(\mathbf{r})$ and $\tilde{\Lambda}^{\dagger}(\mathbf{r})$.  

Equation (\ref{Eq:LambdaTildeHamiltonian}) is an exact reformulation of Eq.\ (\ref{Eq:FieldHamiltonian}); note, however, that $\tilde{\Lambda}(\mathbf{r})$ cannot be straightforwardly expanded in terms of exactly bosonic quasiparticle operators (see Section \ref{SubSec:SecondOrderEquilibrium}), and formulating the many-body Hamiltonian in terms of bosonic quasiparticle operators can be of great utility in determining, for example, energy spectra to high order in a consistent fashion \cite{Morgan2000}.  It is relatively straightforward to determine an equivalent formulation to Eq.\ (\ref{Eq:LambdaTildeHamiltonian}) in terms of $\hat{\Lambda}_{c}(\mathbf{r})$ [Eq.\ (\ref{Eq:LambdaExact})], although this introduces square-root number-operator terms $\sqrt{\hat{N}_{c}}$, which can be awkward to deal with.  

As, in the steady state, the highest-order Hamiltonian considered in this paper is effectively only second-order in the number-conserving fluctuation operators $\tilde{\Lambda}(\mathbf{r})$, $\tilde{\Lambda}^{\dagger}(\mathbf{r})$ (see Section \ref{SubSubSec:TimeIndependentThird}), in the present context such considerations can largely be avoided.

\subsection{Reduction to a third-order Hamiltonian}
\label{SubSec:Reduction}

\subsubsection{Expansion of the condensate number operator}
\label{SubSubSec:Expansion}

If the system is in a number eigenstate of total particle number $N$, the number fluctuations of the condensate and non-condensate components must be equal and opposite.  Formally, 
\begin{equation}
\begin{split}
\hat{N}_{c} - N_{c} =& \int d\mathbf{r}\langle \delta\hat{\Psi}^{\dagger}(\mathbf{r}) \delta\hat{\Psi}(\mathbf{r})\rangle - \int d\mathbf{r} \delta\hat{\Psi}^{\dagger}(\mathbf{r}) \delta\hat{\Psi}(\mathbf{r})\\
=& \int d\mathbf{r}
\left\langle \tilde{\Lambda}^{\dagger}(\mathbf{r})
\frac{N_{c}}{\hat{N}_{c}}
\tilde{\Lambda}(\mathbf{r})\right\rangle 
- \int d\mathbf{r} \tilde{\Lambda}^{\dagger}(\mathbf{r}) 
\frac{N_{c}}{\hat{N}_{c}}
\tilde{\Lambda}(\mathbf{r}).
\end{split}
\label{Eq:NumberFluctuate}
\end{equation}
To quadratic order in $\tilde{\Lambda}(\mathbf{r})$,
\begin{equation}
\hat{N}_{c} = N_{c} + \int d\mathbf{r}\langle \tilde{\Lambda}^{\dagger}(\mathbf{r})
\tilde{\Lambda}(\mathbf{r})\rangle - \int d\mathbf{r} \tilde{\Lambda}^{\dagger}(\mathbf{r}) 
\tilde{\Lambda}(\mathbf{r})
\label{Eq:QuadraticNumberFluctuate}
\end{equation}
(the first corrections beyond this appear at quartic order and are not considered in this paper).  To zeroth order $\hat{N}_{c}=N_{c}$.

We now apply Eq.\ (\ref{Eq:QuadraticNumberFluctuate}) to Eq.\ (\ref{Eq:LambdaTildeHamiltonian}), keeping only terms of up to third order in the fluctuation terms.  Pragmatically, this is equivalent to immediately abandoning the fourth-order term in Eq.\ (\ref{Eq:LambdaTildeHamiltonian}), substituting $N_{c}$ for $\hat{N}_{c}$ in the second- and third-order terms, and substituting Eq.\ (\ref{Eq:QuadraticNumberFluctuate}) into the zeroth- and first-order terms.  This then produces:
\begin{equation}
\begin{split}
\hat{H}_{3} =& N_{c}
\int d\mathbf{r} \phi^{*}(\mathbf{r})
\left[
H_{\mbox{\scriptsize sp}}(\mathbf{r})
+
\frac{\tilde{U}}{2}
 |\phi(\mathbf{r})|^{2}
\right]
\phi(\mathbf{r})
\\&+
\sqrt{N_{c}}
\int d\mathbf{r} 
\left\{
\phi^{*}(\mathbf{r})
\left[
H_{\mbox{\scriptsize sp}}(\mathbf{r})
+
\tilde{U}
 |\phi(\mathbf{r})|^{2}
 \right]
\tilde{\Lambda}(\mathbf{r})
+\mbox{H.c.}
\right\}
\\&+
\int d\mathbf{r} 
\tilde{\Lambda}^{\dagger}(\mathbf{r})
\left[
H_{\mbox{\scriptsize sp}}(\mathbf{r})
+
2\tilde{U}
 |\phi(\mathbf{r})|^{2}
 \right]
\tilde{\Lambda}(\mathbf{r})
\\&+
\frac{\tilde{U}}{2}\int d\mathbf{r} 
\left[
\phi^{*}(\mathbf{r})^{2}\tilde{\Lambda}(\mathbf{r})^{2} +
\mbox{H.c.}
\right]-
\frac{\tilde{U}}{2}
\int d\mathbf{r}
 |\phi(\mathbf{r})|^{4}
\\& +
\int d\mathbf{r'}
\left[
\langle \tilde{\Lambda}^{\dagger}(\mathbf{r'})
\tilde{\Lambda}(\mathbf{r'})\rangle - \tilde{\Lambda}^{\dagger}(\mathbf{r'}) 
\tilde{\Lambda}(\mathbf{r'})
\right]
\\&\times
\int d\mathbf{r}
\phi^{*}(\mathbf{r})
\left[
H_{\mbox{\scriptsize sp}}(\mathbf{r})
+
\tilde{U}
 |\phi(\mathbf{r})|^{2}
 \right]
\phi(\mathbf{r})
\\&+
\frac{\tilde{U}}{\sqrt{N_{c}}}
\int d\mathbf{r} 
\left[
\phi^{*}(\mathbf{r})\tilde{\Lambda}^{\dagger}(\mathbf{r})
\tilde{\Lambda}(\mathbf{r})^{2}
+\mbox{H.c.}
\right]
\\ &-
 \frac{\tilde{U}}{\sqrt{N_{c}}}
\int d\mathbf{r}
\left[
\phi^{*}(\mathbf{r})|\phi(\mathbf{r})|^{2}\tilde{\Lambda}(\mathbf{r}) 
+\mbox{H.c.}
\right]
\\&+
\frac{\tilde{U}}{\sqrt{N_{c}}}
\iint d\mathbf{r}d\mathbf{r'} 
\Bigl\{
\phi^{*}(\mathbf{r})|\phi(\mathbf{r})|^{2} 
\\&\times
\left[
\langle \tilde{\Lambda}^{\dagger}(\mathbf{r'})
\tilde{\Lambda}(\mathbf{r'})\rangle - \tilde{\Lambda}^{\dagger}(\mathbf{r'}) 
\tilde{\Lambda}(\mathbf{r'})
\right]
\tilde{\Lambda}(\mathbf{r}) 
+\mbox{H.c.}
\Bigr\},
\end{split}
\label{Eq:LambdaTildeHamiltonianThirdOrder}
\end{equation}
where the terms have been arranged in descending order of powers of $\sqrt{N_{c}}$.

\subsubsection{Gaussian approximation of the fluctuation terms}
\label{SubSubSec:Gaussian}

In the present context, a Gaussian approximation means that all expectation values of products of the fluctuation operators
$\tilde{\Lambda}(\mathbf{r})$, $\tilde{\Lambda}^{\dagger}(\mathbf{r})$ are either zero (for odd products), or expressable in terms of products of pair-averages \cite{Bach2005}.  Pragmatically, in the equation of motion derived for $\tilde{\Lambda}(\mathbf{r})$, which we determine up to quadratic order in the fluctuation operators, all quadratic products of $\tilde{\Lambda}(\mathbf{r})$ and $\tilde{\Lambda}^{\dagger}(\mathbf{r})$ must be replaced by their expectation values.  Doing this guarantees, for example, that a consistent Gaussian approximant to the equation of motion for the pair-average $\langle \tilde{\Lambda}^{\dagger}(\mathbf{r})\tilde{\Lambda}(\mathbf{r'})\rangle$ is deduced directly from
\begin{equation}
\begin{split}
\frac{d}{dt}\langle \tilde{\Lambda}^{\dagger}(\mathbf{r})\tilde{\Lambda}(\mathbf{r'})\rangle
=& \left\langle
\left[\frac{d}{dt} \tilde{\Lambda}^{\dagger}(\mathbf{r})\right]\tilde{\Lambda}(\mathbf{r'})
\right\rangle
\\&+
 \left\langle
\tilde{\Lambda}^{\dagger}(\mathbf{r})
\left[\frac{d}{dt}\tilde{\Lambda}(\mathbf{r'})\right]
\right\rangle,
\end{split}
\label{Eq:NormalAverageEOM}
\end{equation}
without there being any subsequent need for expansion of  expectation values of higher-order products of the fluctuation operators in terms of pair-averages \cite{Kohler2002}.

An appropriate approximate Hamiltonian $\hat{H}_{3}$ consistent with this desired level of approximation in the equations of motion, should thus be such that the \textit{commutator\/} 
$[\tilde{\Lambda}(\mathbf{r}),\hat{H}_{3}]$ produces terms contributing to the equation of motion for $\tilde{\Lambda}(\mathbf{r})$ in the desired form. This means either scalar terms to zeroth order in the fluctuation operators, first-order operator-valued terms, or scalar second-order terms in the form of pair-averages.  From Eq.\ (\ref{Eq:LambdaTildeCommutator}) and Eq.\ (\ref{Eq:QuadraticNumberFluctuate}) it can be seen that, to quadratic order,
\begin{equation}
\begin{split}
[ \tilde{\Lambda}(\mathbf{r}),\tilde{\Lambda}^{\dagger}(\mathbf{r'})] \approx &
Q(\mathbf{r},\mathbf{r'}) 
\left\{
1+
\int d \mathbf{r''}
\frac{
\langle
\tilde{\Lambda}^{\dagger}(\mathbf{r''})
\tilde{\Lambda}(\mathbf{r''})
\rangle}{N_{c}}
\right.
\\&
\left. 
-
\int d \mathbf{r''}
\frac{
\tilde{\Lambda}^{\dagger}(\mathbf{r''})
\tilde{\Lambda}(\mathbf{r''})
}
{N_{c}
}
\right\}
 -
\frac{\hat{\Lambda}^{\dagger}(\mathbf{r'})\hat{\Lambda}(\mathbf{r})}{N_{c}},
\end{split}
\label{Eq:LambdaTildeCommutatorLambdaPreGauss}
\end{equation} 
and that the first corrections appear at quartic order.
To a
Gaussian level of approximation, the operator products are consistently replaced by expectation values. Thus,
\begin{equation}
[ \tilde{\Lambda}(\mathbf{r}),\tilde{\Lambda}^{\dagger}(\mathbf{r'})] \approx
Q(\mathbf{r},\mathbf{r'})  -
\frac{\langle \hat{\Lambda}^{\dagger}(\mathbf{r'})\hat{\Lambda}(\mathbf{r}) \rangle}{N_{c}}.
\label{Eq:LambdaTildeCommutatorLambda}
\end{equation} 
When considering quadratic and cubic terms in the postulated third-order Hamiltonian $\hat{H}_{3}$, this commutator is simplified further to 
\begin{equation}
[\tilde{\Lambda}(\mathbf{r}),\tilde{\Lambda}^{\dagger}(\mathbf{r'})]  = Q(\mathbf{r},\mathbf{r'}),
\label{Eq:LambdaTildeCommutatorZeroth}
\end{equation}
as otherwise cubic and quartic terms appear in the final equation of motion.

Hence, we deduce that the cubic fluctuation operator products appearing in $\hat{H}_{3}$ [Eq. (\ref{Eq:LambdaTildeHamiltonianThirdOrder})] must be expanded into sums of linear operator-valued terms multiplied by pair-averages.   To this degree of approximation, this is accomplished by expressing cubic products as the sum of all possible pair-averages, multiplied by the remaining fluctuation operator.  This is equivalent to a Hartree-Fock factorization, as described, for example, in Ref.\ \cite{Hutchinson2000}.

For example
\begin{equation}
\begin{split}
\tilde{\Lambda}^{\dagger}(\mathbf{r})
\tilde{\Lambda}(\mathbf{r'})
\tilde{\Lambda}(\mathbf{r''})
\approx &
\langle
\tilde{\Lambda}^{\dagger}(\mathbf{r})
\tilde{\Lambda}(\mathbf{r'})
\rangle
\tilde{\Lambda}(\mathbf{r''})
+\langle
\tilde{\Lambda}^{\dagger}(\mathbf{r})
\tilde{\Lambda}(\mathbf{r''})
\rangle
\tilde{\Lambda}(\mathbf{r'})
\\
&+
\langle
\tilde{\Lambda}(\mathbf{r'})
\tilde{\Lambda}(\mathbf{r''})
\rangle
\tilde{\Lambda}^{\dagger}(\mathbf{r}),
\end{split}
\end{equation}
and we deduce that factorising the cubic terms appearing in Eq.\ (\ref{Eq:LambdaTildeHamiltonianThirdOrder}) results in:
\begin{equation}
\begin{split}
\hat{H}_{3} =& N_{c}
\int d\mathbf{r} \phi^{*}(\mathbf{r})
\left[
H_{\mbox{\scriptsize sp}}(\mathbf{r})
+
\frac{\tilde{U}}{2}
 |\phi(\mathbf{r})|^{2}
\right]
\phi(\mathbf{r})
\\&+
\sqrt{N_{c}}
\int d\mathbf{r} 
\left\{
\phi^{*}(\mathbf{r})
\left[
H_{\mbox{\scriptsize sp}}(\mathbf{r})
+
\tilde{U}
 |\phi(\mathbf{r})|^{2}
 \right]
\tilde{\Lambda}(\mathbf{r})
+\mbox{H.c.}
\right\}
\\&+
\int d\mathbf{r} 
\tilde{\Lambda}^{\dagger}(\mathbf{r})
\left[
H_{\mbox{\scriptsize sp}}(\mathbf{r})
+
2\tilde{U}
 |\phi(\mathbf{r})|^{2}
 \right]
\tilde{\Lambda}(\mathbf{r})
\\&+
\frac{\tilde{U}}{2}\int d\mathbf{r} 
\left[
\phi^{*}(\mathbf{r})^{2}\tilde{\Lambda}(\mathbf{r})^{2} +
\mbox{H.c.}
\right]
-
\frac{\tilde{U}}{2}
\int d\mathbf{r}
 |\phi(\mathbf{r})|^{4}
\\&+
\int d\mathbf{r'}
\left[
\langle \tilde{\Lambda}^{\dagger}(\mathbf{r'})
\tilde{\Lambda}(\mathbf{r'})\rangle - \tilde{\Lambda}^{\dagger}(\mathbf{r'}) 
\tilde{\Lambda}(\mathbf{r'})
\right]
\\&\times
\int d\mathbf{r}
\phi^{*}(\mathbf{r})
\left[
H_{\mbox{\scriptsize sp}}(\mathbf{r})
+
\tilde{U}
 |\phi(\mathbf{r})|^{2}
 \right]
\phi(\mathbf{r})
\\&+
\frac{\tilde{U}}{\sqrt{N_{c}}}
\int d\mathbf{r} 
\Bigl\{
\phi^{*}(\mathbf{r})
\\&\times
\left[
2\langle\tilde{\Lambda}^{\dagger}(\mathbf{r})\tilde{\Lambda}(\mathbf{r})\rangle\tilde{\Lambda}(\mathbf{r})
+
\tilde{\Lambda}^{\dagger}(\mathbf{r})\langle\tilde{\Lambda}(\mathbf{r})^{2}\rangle
\right]
+\mbox{H.c.}
\Bigr\}
\\&-
\frac{\tilde{U}}{\sqrt{N_{c}}}
\int d\mathbf{r}
\left[
\phi^{*}(\mathbf{r})|\phi(\mathbf{r})|^{2}\tilde{\Lambda}(\mathbf{r}) 
+
\mbox{H.c.}
\right]
\\&+
\frac{\tilde{U}}{\sqrt{N_{c}}}
\iint d\mathbf{r}d\mathbf{r'} 
\Bigl\{
\phi^{*}(\mathbf{r})|\phi(\mathbf{r})|^{2} 
\\&\times
\left[
\langle
\tilde{\Lambda}^{\dagger}(\mathbf{r'}) 
\tilde{\Lambda}(\mathbf{r})
\rangle
\tilde{\Lambda}(\mathbf{r'}) 
+
\tilde{\Lambda}^{\dagger}(\mathbf{r'}) 
\langle
\tilde{\Lambda}(\mathbf{r'})
\tilde{\Lambda}(\mathbf{r}) 
\rangle
\right]
+\mbox{H.c.}
\Bigr\}.
\end{split}
\label{Eq:LambdaTildeHamiltonianThirdOrderGaussian}
\end{equation}
It is with respect to this third-order Hamiltonian that our second-order equations of motion will be defined.

The factorization procedure is analogous to that used in Hartree-Fock-Bogoliubov methods. As such, it is not generally valid; careful consideration reveals this not to be a serious problem in the present specific context, however.  Hartree-Fock-Bogoliubov factorizations have also been applied in the full binary interaction Hamiltonian to both cubic and quartic products of the fluctuation operators.  Work by Morgan \cite{Morgan2000} revealed that factorization of the cubic products omitted terms which were as large as terms of quartic origin which were retained.  We, however, have already eliminated quartic fluctuation terms from consideration, and in the steady state all cubic terms will also be eliminated (see Section \ref{SubSubSec:TimeIndependentThird}).  If extension of the theory to include higher-order terms is desired, this simplification will need to be revisited.

\section{Equations of motion}
\label{Sec:Equations}

\subsection{General properties of the equations of motion}
\label{SubSec:General}

\subsubsection{Explicit time dependences}
\label{SubSubSec:ExplicitTime}

It is convenient to have expressions describing the explicit time-dependence only of
$\hat{a}_{c}^{\dagger}$ and
$\delta\hat{\Psi}(\mathbf{r})$.

Taking the partial time-derivative of Eq.\ (\ref{Eq:CondensateCreate}), and recalling that the bosonic field operator has no explicit time-dependence, we deduce that
\begin{equation}
i\hbar\frac{\partial \hat{a}_{c}^{\dagger}}{\partial t} = \int d\mathbf{r} \hat{\Psi}^{\dagger}(\mathbf{r})
\left[
i\hbar\frac{\partial}{\partial t} \phi(\mathbf{r})
\right].
\label{Eq:CondensateCreationExplicit}
\end{equation} 
Similarly, taking the partial time-derivative of Eq.\ (\ref{Eq:NonCondensateField}) produces
 \begin{equation}
i\hbar\frac{\partial}{\partial t} \delta\hat{\Psi}(\mathbf{r})= \int d\mathbf{r'} 
\left[
i\hbar\frac{\partial}{\partial t} Q(\mathbf{r},\mathbf{r'})
\right]
\hat{\Psi}(\mathbf{r'}).
\label{Eq:NonCondensateExplicitFirst}
\end{equation}
The condensate mode-function $\phi(\mathbf{r})$ is defined to have unit norm, which directly implies
\begin{equation}
\int d\mathbf{r} 
\left[
\frac{\partial}{\partial t}\phi^{*}(\mathbf{r})
\right]
\phi(\mathbf{r})
=
-\int d\mathbf{r} 
\phi^{*}(\mathbf{r})
\left[\frac{\partial}{\partial t}\phi(\mathbf{r})
\right].
\label{Eq:NormConserve}
\end{equation}
The resulting Eq.\ (\ref{Eq:NormConserve}) can then be substituted into Eq.\ (\ref{Eq:NonCondensateExplicitFirst}), producing
 \begin{equation}
\begin{split}
i\hbar\frac{\partial}{\partial t} \delta\hat{\Psi}(\mathbf{r})=&
 -\hat{a}_{c}\int d\mathbf{r'} 
Q(\mathbf{r},\mathbf{r'})
\left[
i\hbar\frac{\partial}{\partial t}\phi(\mathbf{r'})
\right]
\\&
-\phi(\mathbf{r})\int d\mathbf{r'}
\left[
i\hbar\frac{\partial}{\partial t}\phi^{*}(\mathbf{r'})
\right]
\delta\hat{\Psi}(\mathbf{r'}).
\end{split}
\label{Eq:NonCondensateExplicit}
\end{equation}

In Eq.\ (\ref{Eq:CondensateCreationExplicit}) and Eq.\ (\ref{Eq:NonCondensateExplicit}), we have the final forms of the desired expressions.

\subsubsection{Condensate number}
\label{SubSubSec:CondensateNumber}

The general equation of motion for the condensate number operator, $\hat{N}_{c}=\hat{a}_{c}^{\dagger} \hat{a}_{c}$, is 
\begin{equation}
i\hbar\frac{d \hat{N}_{c}}{dt} = [\hat{N}_{c},\hat{H}] + i\hbar\frac{\partial \hat{N}_{c}}{\partial t},
\label{Eq:CondensateNumberGeneral}
\end{equation}
from which the dynamics of $N_{c}=\langle \hat{N}_{c} \rangle$ are deduced by taking the expectation value.  

We  first consider the explicit time dependence in isolation. 
Substituting Eq.\ (\ref{Eq:CondensateCreationExplicit}) and its Hermitian conjugate into
$
\partial \hat{N}_{c}/\partial t =
(\partial \hat{a}_{c}^{\dagger}/\partial t  )\hat{a}_{c}+
\hat{a}_{c}^{\dagger}( \partial \hat{a}_{c}/\partial t)
$
produces
\begin{equation}
\begin{split}
i\hbar \frac{\partial \hat{N}_{c}}{\partial t} = &
\int d\mathbf{r}
\left[
\hat{N}_{c}\phi^{*}(\mathbf{r}) + \sqrt{N_{c}}\tilde{\Lambda}^{\dagger}(\mathbf{r})
\right]i\hbar \frac{\partial}{\partial t}\phi(\mathbf{r})
\\ & +
\int d\mathbf{r}
i\hbar\frac{\partial}{\partial t}\phi^{*}(\mathbf{r})\left[
\hat{N}_{c}\phi(\mathbf{r}) + \sqrt{N_{c}}\tilde{\Lambda}(\mathbf{r})
\right].
\end{split}
\label{Eq:CondensateNumberOperatorExplicitFluctuate}
\end{equation}
Substituting in Eq.\ (\ref{Eq:NormConserve}),
we simplify Eq.\ (\ref{Eq:CondensateNumberOperatorExplicitFluctuate}) to
\begin{equation}
\begin{split}
i\hbar\frac{\partial \hat{N}_{c}}{\partial t} = &
\sqrt{N_{c}}\int d\mathbf{r}
\tilde{\Lambda}^{\dagger}(\mathbf{r})
\left[
i\hbar\frac{\partial}{\partial t}\phi(\mathbf{r})
\right]
\\&
+
\sqrt{N_{c}}\int d\mathbf{r}
\left[
i\hbar\frac{\partial}{\partial t}\phi^{*}(\mathbf{r})
\right]
\tilde{\Lambda}(\mathbf{r}).
\end{split}
\label{Eq:CondensateNumberOperatorExplicit}
\end{equation}
Equation (\ref{Eq:CondensateNumberOperatorExplicit}) is entirely composed of linear fluctuation operator terms. Hence, there is no explicit time dependence to the condensate number, i.e.,
\begin{equation}
i\hbar\frac{\partial N_{c}}{\partial t} = 
\left\langle 
i\hbar \frac{\partial \hat{N}_{c}}{\partial t}
\right\rangle = 
0.
\label{Eq:CondensateNumberExplicitZero}
\end{equation}
Therefore, to all orders, the entire time-dependence of the condensate number follows from the (implicit) commutator term of Eq.\ (\ref{Eq:CondensateNumberGeneral}):
\begin{equation}
i\hbar\frac{d N_{c}}{dt} = \langle[\hat{N}_{c},\hat{H}]\rangle.
\label{Eq:CondensateNumberDirect}
\end{equation}
In principle this can be determined directly from
the appropriate form of the Hamiltonian $\hat{H}$.  If one is in any case determining the time-evolution of the individual fluctuation operators $\tilde{\Lambda}(\mathbf{r})$, $\tilde{\Lambda}^{\dagger}(\mathbf{r})$, it is generally more convenient to note from Eq.\ (\ref{Eq:QuadraticNumberFluctuate}) that $N_{c}=N-\int d\mathbf{r} \langle \tilde{\Lambda}^{\dagger}(\mathbf{r})\tilde{\Lambda}(\mathbf{r}) \rangle$ to quadratic order, and therefore that
\begin{equation}
\begin{split}
i\hbar\frac{d N_{c}}{dt} =& 
-\int d\mathbf{r}
\left\langle
\tilde{\Lambda}^{\dagger}(\mathbf{r})
\left[
i\hbar\frac{d}{dt}\tilde{\Lambda}(\mathbf{r})
\right]
\right\rangle
\\&
-\int d\mathbf{r}
\left\langle
\left[
i\hbar\frac{d}{dt}
\tilde{\Lambda}^{\dagger}(\mathbf{r})
\right]
\tilde{\Lambda}(\mathbf{r})
\right\rangle, 
\end{split}
\label{Eq:CondensateNumberIndirect}
\end{equation}
to the (quadratic) order considered here.  

\subsubsection{Fluctuation operator}
\label{SubSubSec:FluctuationOperator}

We now consider the dynamics of the number-conserving fluctuation operator $\tilde{\Lambda}(\mathbf{r})$ directly. In general, the Heisenberg time-evolution of the fluctuation operator is given by
\begin{equation}
i\hbar\frac{d}{dt} \tilde{\Lambda}(\mathbf{r}) = 
[
\tilde{\Lambda}(\mathbf{r}),\hat{H}
]
+ i\hbar\frac{\partial}{\partial t} \tilde{\Lambda}(\mathbf{r}).
\label{Eq:FluctuationTotalTimeDependence} 
\end{equation}
We again initially
consider the explicit time-dependence of Eq.\ (\ref{Eq:FluctuationTotalTimeDependence}), which, from the definition of the fluctuation operator given by Eq.\ (\ref{Eq:LambdaTilde}), yields
\begin{equation}
\begin{split}
 i\hbar\frac{\partial}{\partial t} \tilde{\Lambda}(\mathbf{r})= &
-i\hbar\frac{\partial N_{c}}{\partial t}
\frac{1}{2N_{c}\sqrt{N_{c}}}\hat{a}_{c}^{\dagger}\delta\hat{\Psi}(\mathbf{r})\\
&+\frac{1}{\sqrt{N_{c}}} i\hbar\frac{\partial \hat{a}_{c}^{\dagger}}{\partial t}\delta\hat{\Psi}(\mathbf{r})\\
&+\frac{1}{\sqrt{N_{c}}}\hat{a}_{c}^{\dagger} i\hbar\frac{\partial }{\partial t}\delta\hat{\Psi}(\mathbf{r}).
\end{split}
\label{Eq:LambdaTildeExplicit}
\end{equation}
As there is no explicit time-dependence to $N_{c}$ [Eq.\ (\ref{Eq:CondensateNumberExplicitZero})], the first line of Eq.\ (\ref{Eq:LambdaTildeExplicit}) can be eliminated. After
substituting in Eq.\ (\ref{Eq:CondensateCreationExplicit}) and Eq.\ (\ref{Eq:NonCondensateExplicit}), 
what remains can be expanded in terms of fluctuation and condensate-number operators:
\begin{equation}
\begin{split}
i\hbar \frac{\partial}{\partial t} \tilde{\Lambda}(\mathbf{r})= &
-\frac{\hat{N}_{c}}{\sqrt{N_{c}}}
\int d\mathbf{r'} Q(\mathbf{r},\mathbf{r'})
\left[
i\hbar\frac{\partial}{\partial t} \phi(\mathbf{r'})
\right]
\\&
-\phi(\mathbf{r})\int d\mathbf{r'}
\left[
i\hbar\frac{\partial }{\partial t}\phi^{*}(\mathbf{r'})
\right]
\tilde{\Lambda}(\mathbf{r'})
\\&
+\int d\mathbf{r'}\phi^{*}(\mathbf{r'})
\left[
i\hbar\frac{\partial}{\partial t}\phi(\mathbf{r'})
\right]
\tilde{\Lambda}(\mathbf{r})
\\&
+\frac{1}{\sqrt{N_{c}}} \int d\mathbf{r'} 
\left[
i\hbar\frac{\partial}{\partial t}\phi(\mathbf{r'})
\right]
\tilde{\Lambda}^{\dagger}(\mathbf{r'})\frac{N_{c}}{\hat{N}_{c}}\tilde{\Lambda}(\mathbf{r}).
\end{split}
\label{Eq:LambdaTildeExplicitFull}
\end{equation}

Working within the Gaussian approximation described in Section \ref{SubSubSec:Gaussian}, we
retain terms to first-order in the fluctuation operator $\tilde{\Lambda}(\mathbf{r})$, replace second-order terms with their expectation values, and neglect higher-order terms altogether.  Equation (\ref{Eq:LambdaTildeExplicitFull}) then simplifies to
\begin{equation}
\begin{split}
 i\hbar\frac{\partial}{\partial t} \tilde{\Lambda}(\mathbf{r})= &
-\sqrt{N_{c}}
\int d\mathbf{r'} Q(\mathbf{r},\mathbf{r'})
\left[
i\hbar\frac{\partial}{\partial t} \phi(\mathbf{r'})
\right]
\\&
-\phi(\mathbf{r})\int d\mathbf{r'}
\left[
i\hbar\frac{\partial }{\partial t}\phi^{*}(\mathbf{r'})
\right]
\tilde{\Lambda}(\mathbf{r'})
\\&
+\int d\mathbf{r'}\phi^{*}(\mathbf{r'})
\left[
i\hbar\frac{\partial}{\partial t}\phi(\mathbf{r'})
\right]
\tilde{\Lambda}(\mathbf{r})
\\&
+\frac{1}{\sqrt{N_{c}}} \int d\mathbf{r'} 
\left[
i\hbar\frac{\partial}{\partial t}\phi(\mathbf{r'})
\right]
\langle
\tilde{\Lambda}^{\dagger}(\mathbf{r'})\tilde{\Lambda}(\mathbf{r})
\rangle.
\end{split}
\label{Eq:LambdaTildeExplicitGaussian}
\end{equation}

Which of the terms  of Eq.\ (\ref{Eq:LambdaTildeExplicitGaussian}) are subsequently retained depends on the order to which one wishes to carry out a given calculation.  In order to determine the full dynamics to the desired order, we need to know the form of the appropriate approximate Hamiltonian. Sections \ref{SubSec:FirstOrderHamiltonian}, \ref{SubSec:SecondOrderHamiltonian}, and \ref{SubSec:ThirdOrderHamiltonian} deduce such Hamiltonians to  first, second, and third order, respectively, as well as the associated time-evolutions implied by them.

\subsection{Reduced first-order Hamiltonian}
\label{SubSec:FirstOrderHamiltonian}

\subsubsection{Reduction to a first-order Hamiltonian}
\label{SubSubSec:ReductionFirst}

In principle, one can consider a zeroth-order approximation to the Hamiltonian of Eq.\ (\ref{Eq:LambdaTildeHamiltonianThirdOrderGaussian}).  This is obtained by neglecting all fluctuation terms, and yields a classical energy functional
\begin{equation}
H_{0} = N_{c}
\int d\mathbf{r} \phi^{*}(\mathbf{r})
\left[
H_{\mbox{\scriptsize sp}}(\mathbf{r})
+
\frac{\tilde{U}}{2}
 |\phi(\mathbf{r})|^{2}
\right]
\phi(\mathbf{r}).
\label{Eq:EnergyFunctional}
\end{equation}
The lowest order Hamiltonian of real interest to us is linear in the fluctuation operators $\tilde{\Lambda}(\mathbf{r})$, $\tilde{\Lambda}(\mathbf{r})$, which is when it first has a definite operator character.  Dropping all terms of second and third order in the fluctuation operators from Eq.\ (\ref{Eq:LambdaTildeHamiltonianThirdOrderGaussian}) leaves the appropriate first-order form of the Hamiltonian:
\begin{equation}
\begin{split}
\hat{H}_{1} =& N_{c}
\int d\mathbf{r} \phi^{*}(\mathbf{r})
\left[
H_{\mbox{\scriptsize sp}}(\mathbf{r})
+
\frac{\tilde{U}}{2}
 |\phi(\mathbf{r})|^{2}
\right]
\phi(\mathbf{r})
\\
&+\sqrt{N_{c}}
\int d\mathbf{r} 
\left\{
\phi^{*}(\mathbf{r})
\left[
H_{\mbox{\scriptsize sp}}(\mathbf{r})
+
\tilde{U}
 |\phi(\mathbf{r})|^{2}
 \right]
\tilde{\Lambda}(\mathbf{r})
+\mbox{H.c.}\right\}.
\end{split}
\label{Eq:LambdaTildeHamiltonianFirstOrderGaussian}
\end{equation}

\subsubsection{Deduction of the Gross-Pitaevskii equation}
\label{SubSubSec:TimeEvolutionFirst}

As we are using a first-order approximate Hamiltonian to deduce a zeroth-order approximate equation of motion, we combine Eq.\ (\ref{Eq:FluctuationTotalTimeDependence}) with the first line of Eq.\ (\ref{Eq:LambdaTildeExplicitGaussian}) [the other terms are neglected as being of linear or greater order in $\tilde{\Lambda}(\mathbf{r})$], yielding
\begin{equation}
i\hbar\frac{d}{dt} \tilde{\Lambda}(\mathbf{r}) = 
[
\tilde{\Lambda}(\mathbf{r}),\hat{H}_{1}
 ]
-\sqrt{N_{c}}
\int d\mathbf{r'} Q(\mathbf{r},\mathbf{r'})
\left[
i\hbar\frac{\partial}{\partial t} \phi(\mathbf{r'})
\right].
\label{Eq:LambdaTildeZeroth}
\end{equation}
Using the zeroth-order form of the commutator [Eq.\ (\ref{Eq:LambdaTildeCommutatorZeroth})], inserting the first-order Hamiltonian [Eq.\ (\ref{Eq:LambdaTildeHamiltonianFirstOrderGaussian})] into 
Eq.\  (\ref{Eq:LambdaTildeZeroth}) produces
\begin{equation}
\begin{split}
i\hbar\frac{d}{dt} \tilde{\Lambda}(\mathbf{r}) =& \sqrt{N}_{c}
\int d\mathbf{r'} Q(\mathbf{r},\mathbf{r'})
\\&\times
\left[
H_{\mbox{\scriptsize sp}}(\mathbf{r'})
+
\tilde{U}
 |\phi(\mathbf{r'})|^{2}
-i\hbar\frac{\partial}{\partial t} 
\right]
\phi(\mathbf{r'}).
\end{split}
\label{Eq:LambdaTildeZerothFinal}
\end{equation}
Taking the expectation value of Eq.\ (\ref{Eq:LambdaTildeZerothFinal}), and using the fact that
$
\langle d \tilde{\Lambda}(\mathbf{r})/dt \rangle 
\equiv
d
\langle \tilde{\Lambda}(\mathbf{r})\rangle/dt
=0
$,
we get the time-dependent Gross-Pitaevskii equation, in essentially the same manner as Castin and Dum \cite{Castin1998}, with $N_{c}$ taking the place of $N$ ($\tilde{U}= U_{0}N_{c}$):
\begin{equation}
i\hbar\frac{\partial}{\partial t} \phi(\mathbf{r})=\left[
H_{\mbox{\scriptsize sp}}(\mathbf{r})
+
\tilde{U}
 |\phi(\mathbf{r})|^{2}
-\lambda_{0} \right]
\phi(\mathbf{r}),
\label{Eq:SimpleGrossPitaevskii}
\end{equation}
where
\begin{equation}
\lambda_{0} = \int d\mathbf{r}
\phi^{*}(\mathbf{r})
\left[
H_{\mbox{\scriptsize sp}}(\mathbf{r})
+
\tilde{U}
 |\phi(\mathbf{r})|^{2}
 -i\hbar\frac{\partial}{\partial t}\right]\phi (\mathbf{r}).
\label{Eq:lambdaOne}
\end{equation}

By norm conservation [Eq.\ (\ref{Eq:NormConserve})], the scalar value $\lambda_{0}=\lambda_{0}^{*}$,  and is therefore always real. Substituting Eq.\ (\ref{Eq:SimpleGrossPitaevskii}) into Eq.\ (\ref{Eq:LambdaTildeZerothFinal}) then directly implies that
$i\hbar d\tilde{\Lambda}(\mathbf{r})/dt = 0
$, and hence [through Eq.\ (\ref{Eq:CondensateNumberIndirect})] that $d N_{c}/dt=0$, i.e., to this order, there is no time-dependence to the non-condensate component, and hence no change in the number of non-condensate atoms.  

This is to be expected, as we are considering the system dynamics to zeroth order in the fluctuation operators. Thus, to the current order, we are ignoring the fluctuation operators altogether in the equations of motion.

\subsubsection{Time-independent case}
\label{SubSubSec:TimeIndependentFirst}

Assuming $\phi(\mathbf{r})$ to be a steady state with respect to Eq.\ (\ref{Eq:SimpleGrossPitaevskii}) (generally, although not necessarily the lowest energy steady state), one derives the time-independent Gross-Pitaevskii equation
\begin{equation}
\lambda_{0}\phi(\mathbf{r})=\left[
H_{\mbox{\scriptsize sp}}(\mathbf{r})
+
\tilde{U}
 |\phi(\mathbf{r})|^{2}
 \right]
\phi(\mathbf{r}),
\label{Eq:SimpleGrossPitaevskiiTimeIndependent}
\end{equation}
where $\lambda_{0}$ takes the form of a nonlinear eigenvalue, which at this level of approximation can be identified with the chemical potential.  A consequence of this is that the linear terms in the first-order Hamiltonian [Eq.\ (\ref{Eq:LambdaTildeZerothFinal})] can be eliminated, reducing $\hat{H}_{1}$ to the zeroth-order form given in Eq.\ (\ref{Eq:EnergyFunctional}).

\subsection{Reduced second-order Hamiltonian}
\label{SubSec:SecondOrderHamiltonian}

\subsubsection{Reduction to a second-order Hamiltonian}
\label{SubSubSec:ReductionSecond}

Dropping all terms cubic in the fluctuation operators, $\tilde{\Lambda}(\mathbf{r})$ and $\tilde{\Lambda}^{\dagger}(\mathbf{r})$, from Eq.\ (\ref{Eq:LambdaTildeHamiltonianThirdOrderGaussian}) yields
\begin{equation}
\begin{split}
\hat{H}_{2} =& N_{c}
\int d\mathbf{r} \phi^{*}(\mathbf{r})
\left[
H_{\mbox{\scriptsize sp}}(\mathbf{r})
+
\frac{\tilde{U}}{2}
 |\phi(\mathbf{r})|^{2}
\right]
\phi(\mathbf{r})
-\frac{\tilde{U}}{2}
\int d\mathbf{r}
 |\phi(\mathbf{r})|^{4}
\\&+
\iint d\mathbf{r}
d\mathbf{r'}
\langle \tilde{\Lambda}^{\dagger}(\mathbf{r'})
\tilde{\Lambda}(\mathbf{r'})\rangle
\phi^{*}(\mathbf{r})
\left[
H_{\mbox{\scriptsize sp}}(\mathbf{r})
+
\tilde{U}
 |\phi(\mathbf{r})|^{2}
 \right]
\phi(\mathbf{r})
\\
&+\sqrt{N_{c}}
\int d\mathbf{r} 
\left\{
\phi^{*}(\mathbf{r})
\left[
H_{\mbox{\scriptsize sp}}(\mathbf{r})
+
\tilde{U}
 |\phi(\mathbf{r})|^{2}
 \right]
\tilde{\Lambda}(\mathbf{r})
+\mbox{H.c.}
\right\}
\\&
+
\int d\mathbf{r} 
\tilde{\Lambda}^{\dagger}(\mathbf{r})
\left[
H_{\mbox{\scriptsize sp}}(\mathbf{r})
+
2\tilde{U}
 |\phi(\mathbf{r})|^{2}
 \right]
\tilde{\Lambda}(\mathbf{r})
\\&
+
\frac{\tilde{U}}{2}\int d\mathbf{r} 
\left[
\phi^{*}(\mathbf{r})^{2}\tilde{\Lambda}(\mathbf{r})^{2} +
\mbox{H.c.}
\right]
\\& 
-\iint d\mathbf{r}d\mathbf{r'}
\tilde{\Lambda}^{\dagger}(\mathbf{r'}) 
\tilde{\Lambda}(\mathbf{r'})
\phi^{*}(\mathbf{r})
\left[
H_{\mbox{\scriptsize sp}}(\mathbf{r})
+
\tilde{U}
 |\phi(\mathbf{r})|^{2}
 \right]
\phi(\mathbf{r}),
\end{split}
\label{Eq:LambdaTildeHamiltonianSecondOrderGaussian}
\end{equation}
where the terms have been arranged such that all scalar terms come first (including fluctuation operator pair-averages), followed by terms linear in the fluctuation operators, and subsequently by quadratic (non-expectation value) fluctuation operator terms.

\subsubsection{Deduction of the modified Bogoliubov-de Gennes equations}
\label{SubSubSec:TimeEvolutionSecond}

To determine the equation of motion for the number-conserving fluctuation operator $\tilde{\Lambda}(\mathbf{r})$ to linear order, we must include the zeroth- and linear-order terms from Eq.\ (\ref{Eq:LambdaTildeExplicitGaussian}), inserting these and the quadratic Hamiltonian [Eq.\ (\ref{Eq:LambdaTildeHamiltonianSecondOrderGaussian})] into
Eq.\ (\ref{Eq:FluctuationTotalTimeDependence}): 
\begin{equation}
\begin{split}
i\hbar\frac{d}{dt} \tilde{\Lambda}(\mathbf{r}) = &
[
\tilde{\Lambda}(\mathbf{r}),\hat{H}_{2}
 ]
-\sqrt{N_{c}}
\int d\mathbf{r'} Q(\mathbf{r},\mathbf{r'})
\left[
i\hbar\frac{\partial}{\partial t} \phi(\mathbf{r'})
\right]
\\&
-\phi(\mathbf{r})\int d\mathbf{r'}
\left[
i\hbar\frac{\partial }{\partial t}\phi^{*}(\mathbf{r'})
\right]
\tilde{\Lambda}(\mathbf{r'})
\\&
+\int d\mathbf{r'}\phi^{*}(\mathbf{r'})
\left[
i\hbar\frac{\partial}{\partial t}\phi(\mathbf{r'})
\right]
\tilde{\Lambda}(\mathbf{r}).
\end{split}
\label{Eq:EOMLinearFirst}
\end{equation}

We continue to use the zeroth-order form of the commutator [Eq.\ (\ref{Eq:LambdaTildeCommutatorZeroth})], as to this order we may still neglect the quadratic correction.  Applying this to Eq.\ (\ref{Eq:EOMLinearFirst}) yields
\begin{equation}
\begin{split}
i\hbar\frac{d}{dt} \tilde{\Lambda}(\mathbf{r}) =& \sqrt{N}_{c}
\int d\mathbf{r'} Q(\mathbf{r},\mathbf{r'})
\\&\times
\left[
H_{\mbox{\scriptsize sp}}(\mathbf{r'})
+
\tilde{U}
 |\phi(\mathbf{r'})|^{2}
 -i\hbar\frac{\partial}{\partial t}\right]
\phi(\mathbf{r'})
\\
&+
\int d\mathbf{r'} 
Q(\mathbf{r},\mathbf{r'})
\left[
H_{\mbox{\scriptsize sp}}(\mathbf{r'})
+
2\tilde{U}
 |\phi(\mathbf{r'})|^{2}
 \right]
\tilde{\Lambda}(\mathbf{r'})
\\&+
\tilde{U}\int d\mathbf{r'}
Q(\mathbf{r},\mathbf{r'}) 
\tilde{\Lambda}^{\dagger}(\mathbf{r'})
\phi(\mathbf{r'})^{2}
\\
& 
-\phi(\mathbf{r})\int d\mathbf{r'}
\left[
i\hbar\frac{\partial }{\partial t}\phi^{*}(\mathbf{r'})
\right]
\tilde{\Lambda}(\mathbf{r'})
\\&-
\tilde{\Lambda}(\mathbf{r})
\int d\mathbf{r'}
\phi^{*}(\mathbf{r'})
\\&\times
\left[
H_{\mbox{\scriptsize sp}}(\mathbf{r'})
+
\tilde{U}
 |\phi(\mathbf{r'})|^{2}
 -i\hbar\frac{\partial}{\partial t} \right]
\phi(\mathbf{r'}).
\end{split}
\label{Eq:EOMLinearMiddle}
\end{equation}
Taking the expectation value produces the same Gross-Pitaevskii equation [Eq.\ (\ref{Eq:SimpleGrossPitaevskii})] deduced in Section \ref{SubSubSec:TimeEvolutionFirst}.  This is due to the fact that no linear terms not already present in Eq.\ (\ref{Eq:LambdaTildeHamiltonianFirstOrderGaussian}) appear in Eq.\ (\ref{Eq:LambdaTildeHamiltonianSecondOrderGaussian}).  

Equation (\ref{Eq:SimpleGrossPitaevskii}) can be substituted back into Eq.\ (\ref{Eq:EOMLinearMiddle}), simplifying it to
\begin{equation}
\begin{split}
i\hbar\frac{d}{dt}\tilde{\Lambda}(\mathbf{r}) = &
\left[ 
H_{\mbox{\scriptsize sp}}(\mathbf{r}) + \tilde{U}
|\phi(\mathbf{r})|^2 -\lambda_{0} 
\right]
\tilde{\Lambda}(\mathbf{r}) 
\\&
+\tilde{U}
\int d\mathbf{r'} Q (\mathbf{r},\mathbf{r'}) 
|\phi(\mathbf{r'})|^2  
\tilde{\Lambda}(\mathbf{r'}) 
\\&
+\tilde{U} \int d\mathbf{r'} Q( \mathbf{r},\mathbf{r'})
 \phi^2 (\mathbf{r'}) 
\tilde{\Lambda}^{\dagger}(\mathbf{r'}).
\end{split}
\label{Eq:TDBDG1}
\end{equation}
Equation (\ref{Eq:TDBDG1}), together with its Hermitian conjugate, form the Bogoliubov-de Gennes equations \cite{Bogoliubov1947,deGennes1966}, modified slightly by the presence of the orthogonal projectors $Q(\mathbf{r},\mathbf{r'})$.  This is equivalent to the result presented by Gardiner \cite{Gardiner1997} and Castin and Dum \cite{Castin1998}, apart from the use of $N_{c}$ rather than $N$.

The presence of the projectors is due to the fact that the definition of the condensate and non-condensate components [Eq.\ (\ref{Eq:FieldPartition})] explicitly guarantees their orthogonality \cite{Fetter1972}. This is not true with a conventional symmetry-breaking approach.
Note, however, that if one considers a spatially homogeneous condensate density, then
\begin{equation}
\begin{split}
i\hbar\frac{d}{dt}\tilde{\Lambda}(\mathbf{r}) = &
\left[ 
H_{\mbox{\scriptsize sp}}(\mathbf{r}) +2 \tilde{U}
|\phi(\mathbf{r})|^2 -\lambda_{0} 
\right]
\tilde{\Lambda}(\mathbf{r}) 
+\tilde{U}
 \phi^2 (\mathbf{r}) 
\tilde{\Lambda}^{\dagger}(\mathbf{r}). 
\end{split}
\label{Eq:TDBDG1Homogeneous}
\end{equation}
which coincides with the conventional form of the Bogoliubov-de Gennes equations \cite{Bogoliubov1947,deGennes1966}.

\subsubsection{Number evolution}
\label{SubSubSec:NumberSecond}

Substituting Eq.\ (\ref{Eq:TDBDG1}), together with its Hermitian conjugate, into  Eq.\ (\ref{Eq:CondensateNumberIndirect}) yields that the condensate number evolves as
\begin{equation}
i\hbar\frac{d N_{c}}{dt} = \tilde{U}\int d\mathbf{r} 
\left[
\phi^{*}(\mathbf{r})^{2}
\langle \tilde{\Lambda}(\mathbf{r})^{2}\rangle 
-\langle\tilde{\Lambda}^{\dagger}(\mathbf{r})^{2}\rangle
\phi(\mathbf{r})^{2}
\right].
\label{Eq:NumberSecond}
\end{equation}

This equation is composed of terms quadratic in the fluctuation operators, even though we have everywhere else neglected equivalent quadratic terms.  One could argue that these contributions should be consistently neglected as being ``small'' compared to the current (linear) order of interest.  Inconsistencies,  such as the fact that when considering a non-steady-state evolution the modified Bogoliubov-de Gennes equations [Eq.\ (\ref{Eq:TDBDG1})] can imply unconstrained growth of the non-condensate fraction without there being any corresponding effect on the condensate evolution [Eq.\ (\ref{Eq:SimpleGrossPitaevskii})] \cite{Gardiner2000,SGardiner2002,Castin1997}, would then be dismissed as being due to a misguided attempt to use a first-order theory to determine the evolution of a second-order quantity.  It is useful to examine what is going on in a little more detail, however.

Caveats about purely quantum fluctuations aside, finite fluctuations generally imply finite pair averages, as discussed in Section \ref{SubSubSec:Overview} and Section \ref{SubSubSec:Fluctuation}.  Fluctuations assert their existence by having an observable effect, through pair averages (as well as possibly higher-order connected correlation functions) \cite{Gardiner2004}; for example, when determining the quantum depletion \cite{Pethick2002,Pitaevskii2003}.  For a fixed, finite total number of particles there is, therefore,  a consistency problem intrinsic to first-order theories in the fluctuation operators.  In Section \ref{SubSec:ThirdOrderHamiltonian} it will be shown that, while extending the theory to second-order modifies the equation of motion for $\phi(\mathbf{r})$, it does not change the form of the equations of motion for $\tilde{\Lambda}(\textbf{r})$ [Eq.\ (\ref{Eq:TDBDG1})], and hence $N_{c}$ [Eq.\ (\ref{Eq:NumberSecond})].  As will be described in Section \ref{SubSubSec:Infinite}, in an infinite particle limit ($N_{c}\rightarrow\infty$ while $N_{t}$ remains finite) the first-order equations of motion of Section \ref{SubSec:SecondOrderHamiltonian} are recovered from the second-order equations of motion.  Strictly speaking, this limit is required for the first-order equations of motion to be meaningful.

When allowing $N_{c}$ (and hence $N$) to tend to infinity while requiring $N_{t}$ to remain finite, it is preferable to consider the time evolutions of $N_{t}$ and $N_{c}/N$, which are both finite.  Instead of Eq.\ (\ref{Eq:NumberSecond}), we then  have 
\begin{equation}
i\hbar\frac{d }{dt} \left(\frac{N_{c}}{N}\right)= \tilde{U}\int d\mathbf{r} 
\left[
\phi^{*}(\mathbf{r})^{2}
\frac{\langle \tilde{\Lambda}(\mathbf{r})^{2}\rangle}{N} 
-\frac{\langle\tilde{\Lambda}^{\dagger}(\mathbf{r})^{2}\rangle}{N}
\phi(\mathbf{r})^{2}
\right],
\label{Eq:NumberSecondOverN}
\end{equation}
which tends to zero on the right-hand-side as $N\rightarrow\infty$, and \begin{equation}
i\hbar\frac{d N_{t}}{dt} = \tilde{U}\int d\mathbf{r} 
\left[
\langle\tilde{\Lambda}^{\dagger}(\mathbf{r})^{2}\rangle
\phi(\mathbf{r})^{2}
-\phi^{*}(\mathbf{r})^{2}
\langle \tilde{\Lambda}(\mathbf{r})^{2}\rangle 
\right],
\label{Eq:NumberSecondThermal}
\end{equation}
which does not.  Within this limit it is perfectly legitimate to use Eq.\ (\ref{Eq:NumberSecondThermal}) to determine the time-evolution of the number of non-condensate particles; change in $N_{t}$ then corresponds to an infinitesimal fractional change in $N_{c}$. Conversely, $N_{c}/N = 1 - N_{t}/N \rightarrow 1$, and so the time-derivative in Eq.\  (\ref{Eq:NumberSecondOverN}) must logically be zero.

Such a limit is not appropriate for the desired self-consistent propagation of condensate and non-condensate dynamics with a fixed, finite total particle number $N$, for which the full second-order  treatment described in Section \ref{SubSec:ThirdOrderHamiltonian} is necessary.

\subsubsection{Time-independent case}
\label{SubSubSec:TimeIndependentSecond}

As in Section \ref{SubSubSec:TimeIndependentFirst}, we substitute the time-independent Gross-Pitaevskii equation [Eq.\ (\ref{Eq:SimpleGrossPitaevskiiTimeIndependent})] into the second-order Hamiltonian [Eq.\ (\ref{Eq:LambdaTildeHamiltonianSecondOrderGaussian})], and eliminate the same linear terms.  This yields a form of the Hamiltonian,
\begin{equation}
\begin{split}
\hat{H}_{2} =& N_{c}
\int d\mathbf{r} \phi^{*}(\mathbf{r})
\left[
H_{\mbox{\scriptsize sp}}(\mathbf{r})
+
\frac{\tilde{U}}{2}
 |\phi(\mathbf{r})|^{2}
\right]
\phi(\mathbf{r})
\\&+
\lambda_{0}\int d\mathbf{r}
\langle \tilde{\Lambda}^{\dagger}(\mathbf{r})
\tilde{\Lambda}(\mathbf{r})\rangle
-
\frac{\tilde{U}}{2}
\int d\mathbf{r}
 |\phi(\mathbf{r})|^{4}
\\
&+
\int d\mathbf{r} 
\tilde{\Lambda}^{\dagger}(\mathbf{r})
\left[
H_{\mbox{\scriptsize sp}}(\mathbf{r})
+
2\tilde{U}
 |\phi(\mathbf{r})|^{2}
-\lambda_{0} \right]
\tilde{\Lambda}(\mathbf{r})
\\&+
\frac{\tilde{U}}{2}\int d\mathbf{r} 
\left[
\phi^{*}(\mathbf{r})^{2}\tilde{\Lambda}(\mathbf{r})^{2} +
\mbox{H.c.}
\right],
\end{split}
\label{Eq:LambdaTildeHamiltonianSecondOrderGaussianTimeIndependent}
\end{equation}
equivalent to that deduced in a number-conserving fashion by Gardiner \cite{Gardiner1997}.

\subsection{Properties of the third-order Hamiltonian}
\label{SubSec:ThirdOrderHamiltonian}

\subsubsection{Gaussian form of the third-order Hamiltonian}
\label{SubSubSec:GaussianThird}
The appropriate Gaussian third-order form of the Hamiltonian is exactly as given in Eq.\ (\ref{Eq:LambdaTildeHamiltonianThirdOrderGaussian}). As in Eq.\ (\ref{Eq:LambdaTildeHamiltonianSecondOrderGaussian}), it is convenient to rearrange the equation such that all scalar terms come first (including fluctuation operator pair-averages), followed by terms linear in the fluctuation operators (including those multiplied by fluctuation operator pair-averages), and subsequently by quadratic fluctuation operator terms:
\begin{equation}
\begin{split}
\hat{H}_{3} =& N_{c}
\int d\mathbf{r} \phi^{*}(\mathbf{r})
\left[
H_{\mbox{\scriptsize sp}}(\mathbf{r})
+
\frac{\tilde{U}}{2}
 |\phi(\mathbf{r})|^{2}
\right]
\phi(\mathbf{r})
-\frac{\tilde{U}}{2}
\int d\mathbf{r}
 |\phi(\mathbf{r})|^{4}
\\&+
\iint d\mathbf{r}d\mathbf{r'}
\langle \tilde{\Lambda}^{\dagger}(\mathbf{r'})
\tilde{\Lambda}(\mathbf{r'})\rangle
\phi^{*}(\mathbf{r})
\left[
H_{\mbox{\scriptsize sp}}(\mathbf{r})+
\tilde{U}
 |\phi(\mathbf{r})|^{2}
 \right]
\phi(\mathbf{r})
\\
&+\sqrt{N_{c}}
\int d\mathbf{r} 
\left\{
\phi^{*}(\mathbf{r})
\left[
H_{\mbox{\scriptsize sp}}(\mathbf{r})
+
\tilde{U}
 |\phi(\mathbf{r})|^{2}
 \right]
\tilde{\Lambda}(\mathbf{r})
+\mbox{H.c.}
\right\}
\\
&+\frac{\tilde{U}}{\sqrt{N_{c}}}
\int d\mathbf{r} 
\left[
2\phi^{*}(\mathbf{r})\langle\tilde{\Lambda}^{\dagger}(\mathbf{r})\tilde{\Lambda}(\mathbf{r})\rangle
\tilde{\Lambda}(\mathbf{r})
+\mbox{H.c.}
\right]
\\&
+
\frac{\tilde{U}}{\sqrt{N_{c}}}
\int d\mathbf{r} 
\left[
\langle\tilde{\Lambda}^{\dagger}(\mathbf{r})^{2}\rangle\phi(\mathbf{r})
\tilde{\Lambda}(\mathbf{r})
+\mbox{H.c.}
\right]
\\
&
-\frac{\tilde{U}}{\sqrt{N_{c}}}
\iint d\mathbf{r}d\mathbf{r'} 
\left\{
|\phi(\mathbf{r})|^{2} 
\left[
\phi^{*}(\mathbf{r})
\langle
\tilde{\Lambda}^{\dagger}(\mathbf{r'}) 
\tilde{\Lambda}(\mathbf{r})
\rangle
\right.
\right.
\\&
\left.
\left.
+\langle
\tilde{\Lambda}^{\dagger}(\mathbf{r'})
\tilde{\Lambda}^{\dagger}(\mathbf{r}) 
\rangle
\phi(\mathbf{r})
\right]
\tilde{\Lambda}(\mathbf{r'}) 
+\mbox{H.c.} \right\}
\\&-
\frac{\tilde{U}}{\sqrt{N_{c}}}
\int d\mathbf{r}
\left[
\phi^{*}(\mathbf{r})|\phi(\mathbf{r})|^{2}\tilde{\Lambda}(\mathbf{r}) 
+\mbox{H.c.}\right]
\\&+
\int d\mathbf{r} 
\tilde{\Lambda}^{\dagger}(\mathbf{r})
\left[
H_{\mbox{\scriptsize sp}}(\mathbf{r})
+
2\tilde{U}
 |\phi(\mathbf{r})|^{2}
 \right]
\tilde{\Lambda}(\mathbf{r})
\\&+
\frac{\tilde{U}}{2}\int d\mathbf{r} 
\left[
\phi^{*}(\mathbf{r})^{2}\tilde{\Lambda}(\mathbf{r})^{2} +
\mbox{H.c.}
\right]
\\
& 
-
\iint d\mathbf{r}d\mathbf{r'}
\tilde{\Lambda}^{\dagger}(\mathbf{r'}) 
\tilde{\Lambda}(\mathbf{r'})
\phi^{*}(\mathbf{r})
\left[
H_{\mbox{\scriptsize sp}}(\mathbf{r})
+
\tilde{U}
 |\phi(\mathbf{r})|^{2}
 \right]
\phi(\mathbf{r}).
\end{split}
\label{Eq:LambdaTildeHamiltonianThirdOrderGaussianReordered}
\end{equation}

\subsubsection{Deduction of the generalized Gross-Pitaevskii equation}
\label{SubSubSec:TimeEvolutionThird}

We now  determine the equation of motion for the number-conserving fluctuation operator $\tilde{\Lambda}(\mathbf{r})$, to quadratic order.  We substitute Eq.\ (\ref{Eq:LambdaTildeExplicitGaussian}), in its entirety, and the Gaussian form of the cubic Hamiltonian
[Eq.\ (\ref{Eq:LambdaTildeHamiltonianThirdOrderGaussianReordered})] into
Eq.\ (\ref{Eq:FluctuationTotalTimeDependence}).   The equation of motion can then be written as:
\begin{equation}
\begin{split}
i\hbar\frac{d}{dt} \tilde{\Lambda}(\mathbf{r}) = &
[
\tilde{\Lambda}(\mathbf{r}),\hat{H}_{3}
 ]
-\sqrt{N_{c}}
\int d\mathbf{r'} 
\left[
Q(\mathbf{r},\mathbf{r'})
-
\frac{\langle \tilde{\Lambda}^{\dagger}(\mathbf{r'}) \tilde{\Lambda}(\mathbf{r})\rangle}{N_{c}}
\right]
\\&\times
\left[
i\hbar\frac{\partial}{\partial t} \phi(\mathbf{r'})
\right]
-\phi(\mathbf{r})\int d\mathbf{r'}
\left[
i\hbar\frac{\partial }{\partial t}\phi^{*}(\mathbf{r'})
\right]
\tilde{\Lambda}(\mathbf{r'})
\\&
+\int d\mathbf{r'}\phi^{*}(\mathbf{r'})
\left[
i\hbar\frac{\partial}{\partial t}\phi(\mathbf{r'})
\right]
\tilde{\Lambda}(\mathbf{r}).
\end{split}
\label{Eq:EOMThirdBeginning}
\end{equation}
To produce a consistent second-order equation of motion, we must now include the quadratic correction to the fluctuation operator commutator, using the full form given by Eq.\ (\ref{Eq:LambdaTildeCommutatorLambda}).  This will also produce cubic and quartic corrections, which should be consistently neglected.  Effectively this means that we use the full form of the commutator when determining the time-dependence due to terms of Eq.\ (\ref{Eq:LambdaTildeHamiltonianThirdOrderGaussianReordered}) that are linear in the fluctuation operators.  Otherwise, the zeroth-order form [Eq.\ (\ref{Eq:LambdaTildeCommutatorZeroth})] will suffice.

Doing this produces, subsequent to some rearrangement,
\begin{equation}
\begin{split}
i\hbar\frac{d}{dt} \tilde{\Lambda}(\mathbf{r}) =
& \sqrt{N}_{c}
\int d\mathbf{r'} 
Q(\mathbf{r},\mathbf{r'})
\Biggl(\Biggl\{
H_{\mbox{\scriptsize sp}}(\mathbf{r'})
\\&+
\tilde{U}
\left[
\left(
1-\frac{1}{N_{c}}
\right)
|\phi(\mathbf{r'})|^{2}
+
2\frac{
\langle
\tilde{\Lambda}^{\dagger}(\mathbf{r'})\tilde{\Lambda}(\mathbf{r'})
\rangle 
}{N_{c}}
\right] 
\\&
 -i\hbar\frac{\partial}{\partial t}
\Biggr\}
\phi(\mathbf{r'})
+\tilde{U}\phi^{*}(\mathbf{r'})
\frac{\langle\tilde{\Lambda}(\mathbf{r'})^{2}\rangle}{N_{c}}
\Biggr)
\\&
-\frac{1}{\sqrt{N}_{c}}
\int d\mathbf{r'} 
\Biggl\{
\langle \tilde{\Lambda}^{\dagger}(\mathbf{r'}) \tilde{\Lambda}(\mathbf{r})\rangle
\\&\times
\left[
H_{\mbox{\scriptsize sp}}(\mathbf{r'})
+
2\tilde{U}
 |\phi(\mathbf{r'})|^{2}
 -i\hbar\frac{\partial}{\partial t}\right]
\phi(\mathbf{r'})
\\
&+
\tilde{U} 
\phi^{*}(\mathbf{r'})
\langle\tilde{\Lambda}(\mathbf{r'}) 
 \tilde{\Lambda}(\mathbf{r})\rangle
 |\phi(\mathbf{r'})|^{2}
 \Biggr\}
\\
&+
\int d\mathbf{r'} 
Q(\mathbf{r},\mathbf{r'})
\Bigl\{
\Big[
H_{\mbox{\scriptsize sp}}(\mathbf{r'})
\\&+
2\tilde{U}
 |\phi(\mathbf{r'})|^{2}
 \Bigr]
\tilde{\Lambda}(\mathbf{r'})
+
\tilde{U} \tilde{\Lambda}^{\dagger}(\mathbf{r'})
\phi(\mathbf{r'})^{2}\Bigr\}
\\
& 
-\phi(\mathbf{r})\int d\mathbf{r'}
\left[
i\hbar\frac{\partial }{\partial t}\phi^{*}(\mathbf{r'})
\right]
\tilde{\Lambda}(\mathbf{r'})
\\&-
\tilde{\Lambda}(\mathbf{r})
\int d\mathbf{r'}
\phi^{*}(\mathbf{r'})
\\&\times
\left[
H_{\mbox{\scriptsize sp}}(\mathbf{r'})
+
\tilde{U}
 |\phi(\mathbf{r'})|^{2}
 -i\hbar\frac{\partial}{\partial t} \right]
\phi(\mathbf{r'}).
\end{split}
\label{Eq:EOMThirdMiddle}
\end{equation}
As in Section \ref{SubSubSec:TimeEvolutionSecond}, taking the expectation value of this  expression eliminates all the linear fluctuation terms, leaving us with an equation of motion for the condensate mode $\phi(\mathbf{r})$.   Unlike the simple Gross-Pitaevskii equation [Eq.\ (\ref{Eq:SimpleGrossPitaevskii})], this equation of motion couples to normal and anomalous pair-averages of the number-conserving fluctuation operators: 
\begin{widetext}
\begin{equation}
\begin{split}
i\hbar\frac{\partial}{\partial t} \phi(\mathbf{r})=&\left\{
H_{\mbox{\scriptsize sp}}(\mathbf{r})
+
\tilde{U}
\left[
\left(
1-\frac{1}{N_{c}}
\right)
 |\phi(\mathbf{r})|^{2}
+
2
\frac{\langle
\tilde{\Lambda}^{\dagger}(\mathbf{r})\tilde{\Lambda}(\mathbf{r})
\rangle}{N_{c}}
\right]
-\lambda_{2} \right\}
\phi(\mathbf{r})
+
\tilde{U}
\phi^{*}(\mathbf{r})
\frac{\langle
\tilde{\Lambda}(\mathbf{r})^{2}
\rangle}{N_{c}}
\\
 &-
\int d\mathbf{r'}
\left\{
\frac{\langle
\tilde{\Lambda}^{\dagger}(\mathbf{r'})\tilde{\Lambda}(\mathbf{r})
\rangle}{N_{c}}
\left[
H_{\mbox{\scriptsize sp}}(\mathbf{r'})
+
2\tilde{U}
 |\phi(\mathbf{r'})|^{2}
 -i\hbar\frac{\partial}{\partial t}
\right]
\phi(\mathbf{r'})
+
\tilde{U}
\phi^{*}(\mathbf{r'})
 |\phi(\mathbf{r'})|^{2}
\frac{\langle
\tilde{\Lambda}(\mathbf{r'})\tilde{\Lambda}(\mathbf{r})
\rangle}{N_{c}}
\right\},
\end{split}
\label{Eq:GeneralizedGrossPitaevskiiBeginning}
\end{equation}
where the scalar value, $\lambda_{2}$, is given by
\begin{equation}
\lambda_{2} = \int d\mathbf{r}
\phi^{*}(\mathbf{r})
\left\{
H_{\mbox{\scriptsize sp}}(\mathbf{r})
+
\tilde{U}
\left[
\left(
1-\frac{1}{N_{c}}
\right)
 |\phi(\mathbf{r})|^{2}
+
2
\frac{\langle
\tilde{\Lambda}^{\dagger}(\mathbf{r})\tilde{\Lambda}(\mathbf{r})
\rangle}{N_{c}}
\right]
 -i\hbar\frac{\partial}{\partial t}
 \right\}
\phi(\mathbf{r})
+
\tilde{U}
\int d\mathbf{r}
\phi^{*}(\mathbf{r})^{2}
\frac{\langle
\tilde{\Lambda}(\mathbf{r})^{2}
\rangle}{N_{c}}.
\label{Eq:lambdaTwo}
\end{equation}
\end{widetext}
Note that $\lambda_{2}$, unlike $\lambda_{0}$ [Eq.\ (\ref{Eq:lambdaOne})], may be complex.  The first integral, in a similar fashion to $\lambda_{0}$, can be seen to be always real.   This is not necessarily so for the second integral, as can be seen from
\begin{equation}
\lambda_{2}-\lambda_{2}^{*} = \frac{1}{N_{c}}\tilde{U}
\int d\mathbf{r}
\left[
\phi^{*}(\mathbf{r})^{2}
\langle
\tilde{\Lambda}(\mathbf{r})^{2}
\rangle-
\langle
\tilde{\Lambda}^{\dagger}(\mathbf{r})^{2}
\rangle
\phi(\mathbf{r})^{2}
\right].
\label{Eq:lambdaTwoImaginary}
\end{equation}

We can eliminate the time-derivative on the right-hand side of Eq.\ (\ref{Eq:GeneralizedGrossPitaevskiiBeginning}) by iterative resubstitution, keeping only terms of up to the appropriate order.  This is equivalent to substituting in the lower-order equation of motion for $\phi(\mathbf{r})$, i.e., the Gross-Pitaevskii equation [Eq.\ (\ref{Eq:SimpleGrossPitaevskii})].  Doing this produces 
\begin{widetext}
\begin{equation}
\begin{split}
i\hbar\frac{\partial}{\partial t} \phi(\mathbf{r})=&\left\{
H_{\mbox{\scriptsize sp}}(\mathbf{r})
+
\tilde{U}
\left[
\left(
1-\frac{1}{N_{c}}
\right)
 |\phi(\mathbf{r})|^{2}
+
2
\frac{\langle
\tilde{\Lambda}^{\dagger}(\mathbf{r})\tilde{\Lambda}(\mathbf{r})
\rangle}{N_{c}}
\right]
-\lambda_{2} \right\}
\phi(\mathbf{r})
+
\tilde{U}\phi^{*}(\mathbf{r})
\frac{\langle
\tilde{\Lambda}(\mathbf{r})^{2}
\rangle}{N_{c}}
\\
 &-
\tilde{U}
\int d\mathbf{r'}
 |\phi(\mathbf{r'})|^{2}
\left[
\frac{\langle
\tilde{\Lambda}^{\dagger}(\mathbf{r'})\tilde{\Lambda}(\mathbf{r})
\rangle}{N_{c}}
\phi(\mathbf{r'})
+
\phi^{*}(\mathbf{r'})
\frac{\langle
\tilde{\Lambda}(\mathbf{r'})\tilde{\Lambda}(\mathbf{r})
\rangle}{N_{c}}
\right],
\end{split}
\label{Eq:GeneralizedGrossPitaevskiiEnd}
\end{equation}
\end{widetext}
the final form of the generalized Gross Pitaevskii equation (formally, in conjunction with its complex conjugate, to which it is coupled). This is essentially as was used by Morgan \cite{Morgan2004} to explain finite temperature effects on the excitation spectrum measured in the JILA $^{87}$Rb Bose-Einstein condensate experiment \cite{NoteIn}.  
The anomalous average $\langle
\tilde{\Lambda}(\mathbf{r})^{2}
\rangle$
must be appropriately renormalized to avoid ultraviolet divergences \cite{Hutchinson2000,Morgan2000,Rusch1999,Stoof1993,Proukakis1998,Kohler2002,Bijlsma1997}, as is briefly sketched in Appendix \ref{App:Renormalize}. 

In Ref.\ \cite{Castin1998}, Castin and Dum choose to describe the condensate mode in terms proportional to inverse powers of the total particle number, i.e., as
\begin{equation}
\phi(\mathbf{r}) = \phi_{0}(\mathbf{r}) + \frac{1}{\sqrt{N}}\phi_{1}(\mathbf{r}) + \frac{1}{N}\phi_{2}(\mathbf{r}) + \cdots.
\end{equation}
The zeroth-order term $\phi_{0}(\mathbf{r})$ is then propagated by the Gross-Pitaevskii equation [Eq.\ (\ref{Eq:SimpleGrossPitaevskii}), but Castin and Dum formally have $\tilde{U}= U_{0}N_{c}$ replaced by $U_{0}N$].  The first-order term $\phi_{1}(\mathbf{r})$ is equal to zero [as described in Section \ref{SubSubSec:TimeEvolutionSecond}, there are no first-order corrections to the time evolution of $\phi(\mathbf{r})$].  Equations (95) and (96) of Ref. \cite{Castin1998} describe the time-evolution of the component of $\phi_{2}(\mathbf{r})$ orthogonal to $\phi_{0}(\mathbf{r})$, which  are coupled to the motion of $\phi_{0}(\mathbf{r})$ propagated by the Gross-Pitaevskii  equation.  Combining these equations appears to be equivalent to substituting $\phi(\mathbf{r})=\phi_{0}(\mathbf{r})+\phi_{2}(\mathbf{r})/N$ into Eq.\ (\ref{Eq:GeneralizedGrossPitaevskiiEnd}), dropping terms in $\phi_{2}(\mathbf{r})$ considered to be of too high an order, and replacing  $1/N_{c}$ with $1/N$.  We have found it simpler to consider the evolution of a unified $\phi(\mathbf{r})$ directly, and the deduction of Eq.\ (\ref{Eq:GeneralizedGrossPitaevskiiEnd}) is a significant step toward the paper's main result.

Noting that $\int  d\mathbf{r'} Q(\mathbf{r},\mathbf{r'})\langle \tilde{\Lambda}^{\dagger}(\mathbf{r''})\tilde{\Lambda}(\mathbf{r'})\rangle = \langle \tilde{\Lambda}^{\dagger}(\mathbf{r''})\tilde{\Lambda}(\mathbf{r})\rangle$ and that similarly $\int  d\mathbf{r'} Q(\mathbf{r},\mathbf{r'})\langle \tilde{\Lambda}(\mathbf{r''})\tilde{\Lambda}(\mathbf{r'})\rangle = \langle \tilde{\Lambda}(\mathbf{r''})\tilde{\Lambda}(\mathbf{r})\rangle$, we see that substituting Eq.\ (\ref{Eq:GeneralizedGrossPitaevskiiBeginning}) into Eq.\ (\ref{Eq:EOMThirdMiddle}), the equation of motion for $\tilde{\Lambda}(\mathbf{r})$, causes all terms not linear in the fluctuation operators to vanish.  This is basically equivalent to the removal of the zeroth-order terms when deducing the modified Bogoliubov-de Gennes equations in Section \ref{SubSubSec:TimeEvolutionSecond}.  

One can again substitute Eq.\ (\ref{Eq:GeneralizedGrossPitaevskiiEnd}) for $i\hbar\partial \phi(\mathbf{r})/\partial t$ where it appears in what remains of Eq.\ (\ref{Eq:EOMThirdMiddle}), neglecting all higher order terms; note, however, that this is equivalent to substituting in the simple Gross Pitaevskii equation [Eq.\ (\ref{Eq:SimpleGrossPitaevskii})].  This leaves us with the same modified Bogoliubov-de Gennes equations [Eq.\ (\ref{Eq:TDBDG1})] as determined previously, in Section \ref{SubSubSec:TimeEvolutionSecond}.  

The generalized Gross-Pitaevskii equation [Eq.\ (\ref{Eq:GeneralizedGrossPitaevskiiEnd})], together with the modified Bogoliubov-de Gennes equations [Eq.\ (\ref{Eq:TDBDG1})] thus describe the second-order coupled condensate and non-condensate dynamics, respectively.  This states the main result of the paper.

It should be emphasized that the evolution predicted by the modified Bogoliubov-de Gennes equations may be very different if it is coupled to the \textit{generalized \/} Gross-Pitaevskii equation [Eq.\ (\ref{Eq:GeneralizedGrossPitaevskiiEnd})] rather than the simple Gross-Pitaevskii equation [Eq.\ (\ref{Eq:SimpleGrossPitaevskii})].  That this may constitute a more consistent treatment is shown by the fact that, just as there is an action of the condensate normal and anomalous density terms on the time-evolution of the number conserving fluctuation operators [Eq.\ (\ref{Eq:TDBDG1})], there is a corresponding back action of the normal and anomalous pair-averages on the time-evolution of the condensate mode [Eq.\ (\ref{Eq:GeneralizedGrossPitaevskiiEnd})].  

A similar generalized Gross-Pitaevskii equation can be derived within a symmetry-breaking context \cite{Hutchinson2000}, but without the integral term on the second line of Eq.\ (\ref{Eq:GeneralizedGrossPitaevskiiEnd}). 
Before discussing the role of this term, we note that the projectors $Q(\mathbf{r},\mathbf{r'})$ in the modified Bogoliubov-de Gennes equations [Eq.\ (\ref{Eq:TDBDG1})] can be expanded to give
\begin{widetext}
\begin{equation}
\begin{split}
i\hbar\frac{d}{dt}\tilde{\Lambda}(\mathbf{r}) = &
\left[ 
H_{\mbox{\scriptsize sp}}(\mathbf{r}) + 2\tilde{U}
|\phi(\mathbf{r})|^2 -\lambda_{0} 
\right]
\tilde{\Lambda}(\mathbf{r}) 
-
\phi(\mathbf{r})^2
\tilde{\Lambda}^{\dagger}(\mathbf{r}) 
-\tilde{U}
\int d\mathbf{r'} 
|\phi(\mathbf{r'})|^2  
\left[
\phi^{*}(\mathbf{r'})\phi(\mathbf{r}) 
\tilde{\Lambda}(\mathbf{r'}) 
+
\tilde{\Lambda}^{\dagger}(\mathbf{r'})
\phi(\mathbf{r'})\phi(\mathbf{r})
\right].
\end{split}
\label{Eq:TDBDG1rearrange}
\end{equation}
\end{widetext}
Those parts of the integral terms of Eq.\ (\ref{Eq:GeneralizedGrossPitaevskiiEnd}) and Eq.\ (\ref{Eq:TDBDG1rearrange}) enclosed within square brackets are of almost identical form, but with the roles of the condensate mode functions and the fluctuation operators exchanged.  A comparably elegant simplification of notation afforded by use of the projectors in Eq.\ (\ref{Eq:TDBDG1}) is not obvious for Eq.\ (\ref{Eq:GeneralizedGrossPitaevskiiEnd}). The function of the integral terms in Eq.\ (\ref{Eq:TDBDG1rearrange}) and  Eq.\ (\ref{Eq:GeneralizedGrossPitaevskiiEnd}) is equivalent, however --- to ensure that the orthogonality of the condensate and non-condensate components is maintained \cite{Castin1998}. Hence, their explicitly nonlocal form, and the consequence that both integral terms vanish in the limit of a spatially homogeneous condensate density.  

The appearance of such a term at this order is necessarily in conjunction with the coupling of the generalized  Gross-Pitaevskii equation [Eq.\ (\ref{Eq:GeneralizedGrossPitaevskiiEnd})] to the fluctuation operator normal and anomalous densities.  This is unlike the case in Section \ref{SubSec:SecondOrderHamiltonian}, where the result of the simple time-dependent Gross-Pitaevskii equation [Eq.\ (\ref{Eq:SimpleGrossPitaevskii})] feeds into the modified Bogoliubov-de Gennes equations [Eq.\ (\ref{Eq:TDBDG1})], but the Gross-Pitaevskii equation itself evolves in complete isolation.

\subsubsection{Number evolution}
\label{SubSubSec:NumberThird}

As the time-evolution of the number-conserving fluctuation operators, $\tilde{\Lambda}(\mathbf{r})$  and $\tilde{\Lambda}^{\dagger}(\mathbf{r})$, is still given by the modified Bogoliubov-de Gennes equations [Eq.\ (\ref{Eq:TDBDG1})], the condensate-number evolution must still be given by Eq.\ (\ref{Eq:NumberSecond}).  Note, however, from Eq.\ (\ref{Eq:lambdaTwoImaginary}), that the number dynamics can also be cast as
\begin{equation}
\frac{d N_{c}}{dt} =\frac{ \lambda_{2}-\lambda_{2}^{*}}{i\hbar}N_{c}.
\label{Eq:NumberEvolutionThird}
\end{equation}
This has the form of a simple linear differential equation. The (time-dependent) rate of growth or decay of the number of condensate particles is equal to the difference between the creation of pairs of condensate particles in conjunction with the annihilation of pairs of non-condensate particles, and the reverse process.

The significance of this result is that the condensate-number evolution equation directly implied by the third-order Hamiltonian contains no terms of higher than second-order in the number-conserving fluctuation operators, which is consistent with the order of those fluctuation-operator terms appearing in the generalized Gross-Pitaevskii equation.  This is the lowest non-trivial order at which such a consistent description is possible for a finite number of particles \cite{NoteTrivially}.  One might have expected higher-order fluctuation operator terms to be necessary in the non-condensate evolution for a treatment consistent with the generalized Gross-Pitaevskii equation [\ref{Eq:GeneralizedGrossPitaevskiiEnd}].  This is not so;  consistent \textit{number\/} dynamics in fact require that there be no extension to the modified Bogoliubov-De Gennes equations [Eq.\ (\ref{Eq:TDBDG1})].

\subsubsection{Time-independent case}
\label{SubSubSec:TimeIndependentThird}

If we assume a steady state for $\phi(\mathbf{r})$, then the [equivalent to Eq.\ (\ref{Eq:SimpleGrossPitaevskiiTimeIndependent})] time-independent generalized Gross-Pitaevskii equation is given by
\begin{equation}
\begin{split}
\lambda_{2} \phi(\mathbf{r})=&
\Biggl\{
H_{\mbox{\scriptsize sp}}(\mathbf{r})
+
\tilde{U}
\Biggl[
\left(
1-\frac{1}{N_{c}}
\right)
 |\phi(\mathbf{r})|^{2}
\\&
+
2
\frac{\langle
\tilde{\Lambda}^{\dagger}(\mathbf{r})\tilde{\Lambda}(\mathbf{r})
\rangle}{N_{c}}
\Biggr]
\Biggr\}
\phi(\mathbf{r})
+
\phi^{*}(\mathbf{r})
\tilde{U}\frac{\langle
\tilde{\Lambda}(\mathbf{r})^{2}
\rangle}{N_{c}}
\\
 &-
\int d\mathbf{r'}
\Biggl[
\frac{\langle
\tilde{\Lambda}^{\dagger}(\mathbf{r'})\tilde{\Lambda}(\mathbf{r})
\rangle}{N_{c}}
\tilde{U}
 |\phi(\mathbf{r'})|^{2}
\phi(\mathbf{r'})
\\&+
\phi^{*}(\mathbf{r'})
\tilde{U}
 |\phi(\mathbf{r'})|^{2}
\frac{\langle
\tilde{\Lambda}(\mathbf{r'})\tilde{\Lambda}(\mathbf{r})
\rangle}{N_{c}}
\Biggr].
\end{split}
\label{Eq:GeneralizedGrossPitaevskiiTimeIndependent}
\end{equation}
Substituting this back into Eq.\ (\ref{Eq:LambdaTildeHamiltonianThirdOrderGaussianReordered}), all linear and cubic-order terms disappear.  This is analogous to the way all linear-order terms disappeared in the derivation of the second-order time-independent Hamiltonian, and the elimination of these terms leaves us with the same form of time-independent Hamiltonian [Eq.\ (\ref{Eq:LambdaTildeHamiltonianSecondOrderGaussianTimeIndependent})].

\subsubsection{Infinite-particle limit}
\label{SubSubSec:Infinite}

Examination of Eq.\ (\ref{Eq:GeneralizedGrossPitaevskiiEnd}) and Eq.\ (\ref{Eq:TDBDG1rearrange})
 reveals that allowing the number of condensate particles to arbitrarily increase, i.e.,  $N_{c}\rightarrow \infty$, causes all higher-order terms present in the generalized Gross-Pitaevskii equation to vanish, leaving the simple Gross-Pitaevskii equation [Eq.\ (\ref{Eq:SimpleGrossPitaevskii})], whereas the modified Bogoliubov-de Gennes equations are unchanged.  
 
 We thus reduce exactly to the first-order formulae gained using an approximate second-order Hamiltonian [Eq.\ (\ref{Eq:LambdaTildeHamiltonianSecondOrderGaussian})].  When one considers that a treatment using the modified Bogoliubov-de Gennes equations [Eq.\ (\ref{Eq:TDBDG1})] coupled to the simple Gross-Pitaevskii equation [Eq.\ (\ref{Eq:SimpleGrossPitaevskii})] allows for unlimited growth of the non-condensate fraction without there being any effect on the condensate dynamics, it is clear that only in the limit of an infinite number of condensate particles can the dynamics predicted by these equations be strictly correct.

\section{Equilibrium properties}
\label{Sec:Equilibrium}

\subsection{Overview}
Section \ref{SubSec:SecondOrderEquilibrium} recaps the situations described by Gardiner \cite{Gardiner1997} and Castin and Dum \cite{Castin1998}, which, in addition to work by Girardeau and Arnowitt \cite{Girardeau1959,Girardeau1998}, sought to provide a number-conserving equivalent to the symmetry breaking Bogoliubov formalism \cite{Bogoliubov1947,deGennes1966}.  That is, considering the Hamiltonian to second order in the fluctuation terms, or equivalently, equations of motion of up to first order in the fluctuation terms.  In the present context, this is equivalent to assuming the correctness of Eq.\ (\ref{Eq:SimpleGrossPitaevskii}) and Eq.\ (\ref{Eq:TDBDG1}).  Having set context and notation, Section \ref{SubSec:ThirdOrderEquilibrium} considers some of the difficulties in going beyond this level of approximation.

\subsection{Quasiparticle formulation}
\label{SubSec:SecondOrderEquilibrium}

\subsubsection{Two-component representation}
\label{SubSubSec:Spinor}

As the time-evolution of the number-conserving fluctuation operator $\tilde{\Lambda}(\mathbf{r})$ [Eq.\ (\ref{Eq:TDBDG1})] causes it to couple to its Hermitian conjugate, it can be convenient to write the coupled time-evolution equations in a unified two-component form.  Thus
\begin{equation}
i\hbar \frac{d}{dt}
\left(
\begin{array}{c}
\tilde{\Lambda}(\mathbf{r})
\\
\tilde{\Lambda}^{\dagger}(\mathbf{r})
\end{array}
\right)
=
\int d\mathbf{r'}
\mathcal{J}(\mathbf{r},\mathbf{r'})
\left(
\begin{array}{c}
\tilde{\Lambda}(\mathbf{r'})
\\
\tilde{\Lambda}^{\dagger}(\mathbf{r'})
\end{array}
\right),
\label{Eq:TDBDGSpinor1}
\end{equation}
where
\begin{equation}
\mathcal{J}(\mathbf{r},\mathbf{r'})
=
\left(
\begin{array}{cc}
J(\mathbf{r},\mathbf{r'})
&
K(\mathbf{r},\mathbf{r'})
\\
-K^{*}(\mathbf{r},\mathbf{r'})
&
-J^{*}(\mathbf{r},\mathbf{r'})
\end{array}
\right),
\label{Eq:TDBDGSpinor1JCal}
\end{equation}
and the elements of $\mathcal{J}(\mathbf{r},\mathbf{r'})$ are defined by
\begin{align}
\begin{split}
J(\mathbf{r},\mathbf{r'}) = &
\delta(\mathbf{r}-\mathbf{r'})\left[ 
H_{\mbox{\scriptsize sp}}(\mathbf{r'}) + \tilde{U}
|\phi(\mathbf{r'})|^2 -\lambda_{0} 
\right]
\\&
+Q (\mathbf{r},\mathbf{r'}) 
\tilde{U}|\phi(\mathbf{r'})|^2,  
\end{split}
\label{Eq:TDBDGSpinor1J}
\\
K(\mathbf{r},\mathbf{r'}) = &
Q( \mathbf{r},\mathbf{r'})
\tilde{U}  \phi(\mathbf{r'})^{2},  
\label{Eq:TDBDGSpinor1K}
\end{align}
and their complex conjugates.  

As $\int d\mathbf{r'} Q(\mathbf{r},\mathbf{r'})\tilde{\Lambda}(\mathbf{r'})=\tilde{\Lambda}(\mathbf{r})$, we choose to describe the fluctuation operator time-evolution by
\begin{equation}
i\hbar \frac{d}{dt}
\left(
\begin{array}{c}
\tilde{\Lambda}(\mathbf{r})
\\
\tilde{\Lambda}^{\dagger}(\mathbf{r})
\end{array}
\right)
=
\int d\mathbf{r'}
\mathcal{L}(\mathbf{r},\mathbf{r'})
\left(
\begin{array}{c}
\tilde{\Lambda}(\mathbf{r'})
\\
\tilde{\Lambda}^{\dagger}(\mathbf{r'})
\end{array}
\right),
\label{Eq:TDBDGSpinor2}
\end{equation}
where
\begin{equation}
\mathcal{L}(\mathbf{r},\mathbf{r'})
=
\left(
\begin{array}{cc}
L(\mathbf{r},\mathbf{r'})
&
M(\mathbf{r},\mathbf{r'})
\\
-M^{*}(\mathbf{r},\mathbf{r'})
&
-L^{*}(\mathbf{r},\mathbf{r'})
\end{array}
\right),
\label{Eq:TDBDGSpinor2LCal}
\end{equation}
and the elements of $\mathcal{L}(\mathbf{r},\mathbf{r'})$ are defined by
\begin{align}
\begin{split}
L(\mathbf{r},\mathbf{r'}) = &
\delta(\mathbf{r}-\mathbf{r'})\left[ 
H_{\mbox{\scriptsize sp}}(\mathbf{r'}) + \tilde{U}
|\phi(\mathbf{r'})|^2 -\lambda_{0} 
\right]
\label{Eq:TDBDGSpinor2L}
\\&
+\int d\mathbf{r''}Q (\mathbf{r},\mathbf{r''}) 
\tilde{U}|\phi(\mathbf{r''})|^2
Q (\mathbf{r''},\mathbf{r'}),  
\end{split}
\\
M(\mathbf{r},\mathbf{r'}) = &
\int d\mathbf{r''}Q( \mathbf{r},\mathbf{r''})
\tilde{U}  \phi(\mathbf{r''})^{2}  
Q^{*}( \mathbf{r''},\mathbf{r'}).
\label{Eq:TDBDGSpinor2M}
\end{align}
Note that $L(\mathbf{r'},\mathbf{r})=L^{*}(\mathbf{r},\mathbf{r'})$, i.e., $L$ is Hermitian, and thus $\mathcal{L}(\mathbf{r},\mathbf{r'})$ has some symmetry properties which $\mathcal{J}(\mathbf{r},\mathbf{r'})$ does not \cite{Castin1998}.

Inserting the projectors $Q(\mathbf{r},\mathbf{r'})$ into Eq.\ (\ref{Eq:TDBDGSpinor1}) in this way has the useful property that the evolutions predicted by the modified Bogoliubov-de Gennes equations [Eq.\ (\ref{Eq:TDBDG1})] and the simple Gross-Pitaevskii equation [Eq.\ (\ref{Eq:SimpleGrossPitaevskii})] are unified by the application of the operator $\mathcal{L}(\mathbf{r},\mathbf{r'})$ onto an appropriate two-component state.  Thus, replacing $(\tilde{\Lambda}(\mathbf{r'}),
\tilde{\Lambda}^{\dagger}(\mathbf{r'}))$ in Eq.\ (\ref{Eq:TDBDGSpinor2}) with $(\phi(\mathbf{r}),0)$ or $(0,\phi^{*}(\mathbf{r}))$ reduces it to the simple Gross-Pitaevskii equation, or its complex conjugate, respectively \cite{Castin1998}.

\subsubsection{Quasiparticles}
\label{SubSubSec:Quasiparticles}
The spectral decomposition of $\mathcal{L}(\mathbf{r},\mathbf{r'})$, 
\begin{equation}
\begin{split}
\mathcal{L}(\mathbf{r},\mathbf{r'})=&
\sum_{k=1}^{\infty}
\epsilon_{k}
\left(
\begin{array}{c}
u_{k}(\mathbf{r})
\\
v_{k}(\mathbf{r})
\end{array}
\right)
(
u_{k}^{*}(\mathbf{r'}), -v_{k}^{*}(\mathbf{r'})
)
\\&
-
\sum_{k=1}^{\infty}
\epsilon_{k}
\left(
\begin{array}{c}
v_{k}^{*}(\mathbf{r})
\\
u_{k}^{*}(\mathbf{r})
\end{array}
\right)
(
-v_{k}(\mathbf{r'}), u_{k}(\mathbf{r'})
),
\end{split}
\label{Eq:SpectralDecomp}
\end{equation}
the derivation of which is outlined in Appendix \ref{App:Spectral}, provides a useful basis in which to expand the number conserving fluctuation operators:
\begin{equation}
\left(
\begin{array}{c}
\tilde{\Lambda}(\mathbf{r})
\\
\tilde{\Lambda}^{\dagger}(\mathbf{r})
\end{array}
\right) 
=
\sum_{k=1}^{\infty}
\tilde{b}_{k}
\left(
\begin{array}{c}
u_{k}(\mathbf{r})
\\
v_{k}(\mathbf{r})
\end{array}
\right)
+
\sum_{k=1}^{\infty}
\tilde{b}_{k}^{\dagger}
\left(
\begin{array}{c}
v_{k}^{*}(\mathbf{r})
\\
u_{k}^{*}(\mathbf{r})
\end{array}
\right).
\label{Eq:LDecompMain}
\end{equation}

In turn, using the orthonormality relations
\begin{align}
\int d\mathbf{r}
[u_{k'}^{*}(\mathbf{r})u_{k}(\mathbf{r})
-
v_{k'}^{*}(\mathbf{r})v_{k}(\mathbf{r})]
= & \delta_{kk'},\\
\int d\mathbf{r}
[u_{k'}(\mathbf{r})v_{k}(\mathbf{r})
-
v_{k'}(\mathbf{r})u_{k}(\mathbf{r})]
=& 0,
\end{align}
we determine that the operator coefficients are given by
\begin{align}
\tilde{b}_{k} = & \int d\mathbf{r} 
u_{k}^{*}(\mathbf{r})\tilde{\Lambda}(\mathbf{r})
-
v_{k}^{*}(\mathbf{r})\tilde{\Lambda}^{\dagger}(\mathbf{r})
\\
\tilde{b}_{k}^{\dagger} = & \int d\mathbf{r} 
u_{k}(\mathbf{r})\tilde{\Lambda}^{\dagger}(\mathbf{r})
-
v_{k}(\mathbf{r})\tilde{\Lambda}(\mathbf{r})
\end{align}
and that their commutation relations are
\begin{align}
\begin{split}
[\tilde{b}_{k},\tilde{b}_{k'}^{\dagger}] = &
\iint d\mathbf{r} d\mathbf{r'}
[u_{k}^{*}(\mathbf{r})u_{k'}(\mathbf{r'})-v_{k}^{*}(\mathbf{r'})v_{k'}(\mathbf{r})] 
\\ &\times
[\tilde{\Lambda}(\mathbf{r}),\tilde{\Lambda}^{\dagger}(\mathbf{r'})],
\end{split}
\label{Eq:PreQuasiparticleCommutator}
\\
\begin{split}
[\tilde{b}_{k},\tilde{b}_{k'}] = &
\iint d\mathbf{r} d\mathbf{r'}
[u_{k}^{*}(\mathbf{r})v_{k'}^{*}(\mathbf{r'})-v_{k}^{*}(\mathbf{r'})u_{k'}^{*}(\mathbf{r})] 
\\ &\times
[\tilde{\Lambda}(\mathbf{r}),\tilde{\Lambda}^{\dagger}(\mathbf{r'})],
\end{split}
\label{Eq:PreQuasiparticleCommutatorSame}
\end{align}

If we can assume the commutator for the number-conserving fluctuation operators to be reduced to the projector $Q(\mathbf{r},\mathbf{r'})$ [Eq.\ (\ref{Eq:LambdaTildeCommutatorZeroth})], the operator coefficients $\tilde{b}_{k}$, $\tilde{b}_{k}^{\dagger}$ form a bosonic algebra:
\begin{align}
[\tilde{b}_{k},\tilde{b}_{k'}^{\dagger}] = & \delta_{kk'}, 
\label{Eq:QuasiparticleCommutator}\\
[\tilde{b}_{k},\tilde{b}_{k'}] = & 0.
\label{Eq:QuasiparticleCommutatorSame}
\end{align}

The operators $\tilde{b}_{k}^{\dagger}$ and $\tilde{b}_{k}$ are then quasiparticle creation and annihilation operators \cite{Castin1998,Gardiner1997}.

\subsubsection{Reformulation of the Hamiltonian in terms of quasiparticles}

Substituting Eq.\ (\ref{Eq:LDecompMain}) into Eq.\ (\ref{Eq:LambdaTildeHamiltonianSecondOrderGaussianTimeIndependent}) [and making use of 
Eq.\ (\ref{Eq:LRightEig}), Eq.\ (\ref{Eq:MStarRightEig}), Eq.\ (\ref{Eq:LStarRightEig}), Eq.\ (\ref{Eq:MRightEig}), and Eq.\ (\ref{Eq:InnerProduct})] yields
\begin{equation}
\begin{split}
\hat{H}_{2} =& \mathcal{H}
+\sum_{k,k'=1}^{\infty}
\left(
\frac{\epsilon_{k}+\epsilon_{k'}}{2}
\right)
\tilde{b}_{k}^{\dagger}\tilde{b}_{k'}
\int d\mathbf{r}
u_{k}(\mathbf{r})^{*}u_{k'}(\mathbf{r})
\\&-
\sum_{k,k'=1}^{\infty}
\left(
\frac{\epsilon_{k}+\epsilon_{k'}}{2}
\right)
\tilde{b}_{k'}\tilde{b}_{k}^{\dagger}
\int d\mathbf{r}
v_{k'}(\mathbf{r})v_{k}^{*}(\mathbf{r}),
\end{split}
\label{Eq:LambdaTildeHamiltonianSecondOrderGaussianTimeIndependentQuasiparticle}
\end{equation}
where
\begin{equation}
\begin{split}
\mathcal{H} = &
N_{c}
\int d\mathbf{r} \phi^{*}(\mathbf{r})
\left[
H_{\mbox{\scriptsize sp}}(\mathbf{r})
+
\frac{\tilde{U}}{2}
\left(
1-\frac{1}{N_{c}}
\right)
 |\phi(\mathbf{r})|^{2}
\right]
\phi(\mathbf{r})
\\&+
\lambda_{0}
\sum_{k,k'=1}^{\infty}
\langle
\tilde{b}_{k}\tilde{b}_{k'}^{\dagger}
\rangle
\int d\mathbf{r}v_{k}(\mathbf{r})v_{k'}^{*}(\mathbf{r})
\\&+
\lambda_{0}
\sum_{k,k'=1}^{\infty}
\langle
\tilde{b}_{k}^{\dagger}\tilde{b}_{k'}
\rangle
\int d\mathbf{r}u_{k}^{*}(\mathbf{r})u_{k'}(\mathbf{r})
\\&+
\lambda_{0}
\sum_{k,k'=1}^{\infty}
\langle
\tilde{b}_{k}\tilde{b}_{k'}
\rangle
\int d\mathbf{r}v_{k}(\mathbf{r})u_{k'}(\mathbf{r})
\\&+
\lambda_{0}
\sum_{k,k'=1}^{\infty}
\langle
\tilde{b}_{k}^{\dagger}\tilde{b}_{k'}^{\dagger}
\rangle
\int d\mathbf{r}u_{k}^{*}(\mathbf{r})v_{k'}^{*}(\mathbf{r}).
\end{split}
\label{Eq:ExtendedEnergyFunctional}
\end{equation}
Making use of Eq.\ (\ref{Eq:QuasiparticleCommutator}), i.e., assuming the quasiparticle operators to have bosonic commutation relations, reduces Eq.\ (\ref{Eq:LambdaTildeHamiltonianSecondOrderGaussianTimeIndependentQuasiparticle}) to diagonal form \cite{Bogoliubov1947,Gardiner1997,Castin1998}:
\begin{equation}
\hat{H}_{2} =
\mathcal{H}-\sum_{k=1}^{\infty} \epsilon_{k} \int d\mathbf{r} |v_{k}(\mathbf{r})|^{2}
+\sum_{k=1}^{\infty}
\epsilon_{k}
\tilde{b}_{k}^{\dagger}\tilde{b}_{k},
\end{equation}
and assuming a thermal equilibrium state, $\mathcal{H}$ reduces to 
\begin{equation}
\begin{split}
\mathcal{H} = &
N_{c}
\int d\mathbf{r} \phi^{*}(\mathbf{r})
\left[
H_{\mbox{\scriptsize sp}}(\mathbf{r})
+
\frac{\tilde{U}}{2}
\left(
1-\frac{1}{N_{c}}
\right)
 |\phi(\mathbf{r})|^{2}
\right]
\phi(\mathbf{r})
\\&+
\lambda_{0}
\sum_{k=1}^{\infty}
\langle
\tilde{b}_{k}^{\dagger}\tilde{b}_{k}
\rangle
\int d\mathbf{r}
\left[
u_{k}^{*}(\mathbf{r})u_{k}(\mathbf{r})+v_{k}^{*}(\mathbf{r})v_{k}(\mathbf{r})
\right]
\\&+
\lambda_{0}
\sum_{k=1}^{\infty}
\int d\mathbf{r}v_{k}^{*}(\mathbf{r})v_{k}(\mathbf{r}).
\end{split}
\label{Eq:ExtendedEnergyFunctionalEquilibrium}
\end{equation}

This all being so, the quasiparticle populations for a system in thermal equilibrium are given by
$\langle
\tilde{b}_{k}^{\dagger}\tilde{b}_{k}\rangle = [\exp(\{\epsilon_{k}-[\mu-\lambda_{0}]\}/k_{b}T)-1]^{-1}$ \cite{Morgan2004,Mandl1988}, where $\mu$ is the chemical potential, $T$ the temperature, and $k_{B}$ Boltzmann's constant.
Having populated the system appropriately, one can determine the time-evolution of the fluctuation operators from a system initially at equilibrium purely through the mode functions, such that
\begin{equation}
i\hbar \frac{d}{dt}
\left(
\begin{array}{c}
u_{k}(\mathbf{r})
\\
v_{k}(\mathbf{r})
\end{array}
\right)
=
\int d\mathbf{r'}
\mathcal{L}(\mathbf{r},\mathbf{r'})
\left(
\begin{array}{c}
u_{k}(\mathbf{r'})
\\
v_{k}(\mathbf{r'})
\end{array}
\right),
\label{Eq:TDBDGSpinor3}
\end{equation}
and the $\tilde{b}_{k}$, $\tilde{b}_{k}^{\dagger}$ are constant.

\subsection{Further considerations}
\label{SubSec:ThirdOrderEquilibrium}
As has been shown in Section \ref{SubSubSec:TimeIndependentSecond} and Section \ref{SubSubSec:TimeIndependentThird}, $\hat{H}_{3}$ and $\hat{H}_{2}$ have the same form if the system is in equilibrium [Eq.\ (\ref{Eq:LambdaTildeHamiltonianSecondOrderGaussianTimeIndependent})], meaning that Eq.\ (\ref{Eq:LambdaTildeHamiltonianSecondOrderGaussianTimeIndependentQuasiparticle}) is an equally valid reformulation of $\hat{H}_{3}$ in an equilibrium context.  A concern is that use of the more complete formulation of the commutator [Eq.\ (\ref{Eq:LambdaTildeCommutatorLambda})] reveals that the quasiparticle commutation relations are not exactly bosonic [Eq.\ (\ref{Eq:PreQuasiparticleCommutator}) and Eq.\ (\ref{Eq:PreQuasiparticleCommutatorSame})].  If we recall that,  in conjunction with second-order terms, we have always used the simpler form of the commutator, in the context of the present paper this does not seem to be a critical consideration.  Extending this approach  to a consistent higher-order formalism, as is in principle desirable, will present some difficulties, however. 

We could take slightly different operators, defined from $\hat{\Lambda}_{c}(\mathbf{r})$, 
$\hat{\Lambda}_{c}^{\dagger}(\mathbf{r})$  [Eq.\  (\ref{Eq:LambdaExact})],
\begin{equation}
\left(
\begin{array}{c}
\hat{\Lambda}_{c}(\mathbf{r})
\\
\hat{\Lambda}_{c}^{\dagger}(\mathbf{r})
\end{array}
\right) 
=
\sum_{k=1}^{\infty}
\hat{b}_{k}
\left(
\begin{array}{c}
u_{k}(\mathbf{r})
\\
v_{k}(\mathbf{r})
\end{array}
\right)
+
\sum_{k=1}^{\infty}
\hat{b}_{k}^{\dagger}
\left(
\begin{array}{c}
v_{k}^{*}(\mathbf{r})
\\
u_{k}^{*}(\mathbf{r})
\end{array}
\right).
\label{Eq:LDecompExact}
\end{equation}
As a consequence of the commutation relation described in Eq.\ (\ref{Eq:LambdaCommutatorExact}), the commutation relations of $\hat{b}_{k}$ and $\hat{b}_{k}^{\dagger}$ are exactly bosonic, and therefore could potentially better describe the system in terms of a Bose-Einstein distribution.  This would be more in keeping with the spirit of the detailed treatment, making use of second-order perturbation theory, given in Ref.\ \cite{Morgan2000}. 

As described in Section \ref{SubSec:FluctuationExpansion}, $\hat{\Lambda}_{c}(\mathbf{r})$ is not a simple fluctuation operator. In other words, its expectation value is not defined to be exactly equal to zero. The formal development of the equations of motion in Section \ref{Sec:Equations} explicitly relies on the number-conserving fluctuation operator having expectation value equal to zero.  Hence, using $\hat{\Lambda}_{c}(\mathbf{r})$, which has expectation value only approximately equal to zero, would mean that corrections, particularly at higher order, would have to be carefully calculated and accounted for.  Furthermore, the idea that one is looking at fluctuations about a well defined mean, and that the magnitude of these fluctuations is observed, in the first instance, through pair expectation values, possibly to be followed by a well-defined hierarchy of connected correlation functions or cumulants, is muddied.  In other words, an imprecision in the procedure used to determine the relevant equations of motion is introduced at a basic level.  It therefore does not seem that a demand for such precision is compatible with defining perfectly bosonic quasiparticle operators.  

It should be emphasized that this is not a comment over whether it is more correct to use either $\hat{\Lambda}_{c}(\mathbf{r})$ or $\tilde{\Lambda}(\mathbf{r})$.  Having an expectation value exactly equal to zero, and inducing exactly bosonic quasiparticle commutation relations, are both highly desirable properties for a hypothetical fluctuation operator to have.  Within the formal framework used in this paper, for a finite total number of particles, it seems that one has choose which of these properties is the more important for the purpose at hand.  The emphasis in this paper is on the formal development of the relevant equations of motion.  For this, at least, describing things in terms of $\tilde{\Lambda}(\mathbf{r})$ and $\tilde{\Lambda}^{\dagger}(\mathbf{r})$ seems to be more convenient.  For other purposes, it may be more appropriate to consider a formulation closer to $\hat{\Lambda}_{c}(\mathbf{r})$ and $\hat{\Lambda}_{c}^{\dagger}(\mathbf{r})$ \cite{Morgan2000}.

Good results have been achieved, using the equations of motion developed in this paper, in describing excitations in finite temperature Bose-Einstein condensate by Morgan \cite{Morgan2004}.  We also note that such issues as potentially imperfectly-bosonic quasiparticle operators are largely avoided if the initial system has a negligible non-condensate fraction, even if subsequent dynamics (for example investigations of chaotic dynamics \cite{Gardiner2000,SGardiner2002,Castin1997,Ryu2006,Behinaein2006,Xie2005,Zhang2004,Liu2006,Duffy2004,Salmond2002,Hai2002,Martin2006}) can cause significant depletion \cite{Gardiner2000,SGardiner2002,Castin1997}, hence requiring the kind of self-consistent treatment presented here.

\section{Conclusions}
\label{Sec:Conclusions}

In conclusion, we have shown that a coupled system of equations, the generalized Gross-Pitaevskii equation and the modified Bogoliubov-de Gennes equation are the necessary minimally complete description to imply internally consistent number dynamics for a finite total number of particles.  In other words, dynamics such that only particles lost from the condensate fraction are assumed by the non-condensate fraction, and vice-versa.  Elaboration of the (linear) modified Bogoliubov-de Gennes equations is neither desirable nor necessary, as this would automatically lead to inconsistent number dynamics.  That an approach to second order in the fluctuation operators is necessary is directly implied by elementary statistical considerations; effectively that a finite fluctuation directly implies a finite variance, or its equivalent.  Hence, in an infinite particle limit the first-order approach, consisting of the simple Gross-Pitaevskii equation coupled to the modified Bogoliubov-de Gennes equations, is recovered.  It is only in this limit that the dynamics predicted by this system of equations are technically consistent.  A similar form of the approach presented here has been employed \cite{Morgan2003,Morgan2004,Morgan2005} as a key component of an analysis of the observed excitations in finite temperature Bose-Einstein condensates, to good agreement with experiment \cite{Jin1997}.  The formalism presented here is also suitable for the study of dynamically unstable Bose-Einstein condensate dynamics, where, even if the sample is initially at zero temperature, it is possible for a sizable non-condensate fraction to build up over time.

\section*{Acknowledgements}
We thank K. Burnett, J. Cooper, C. W. Gardiner, I. G. Hughes, and M. J. Holland for numerous enlightening discussions over an extended period of time, and the UK EPSRC (grant no.\ EP/D032970/1) and the Royal Society of London for support.

\begin{appendix}

\section{Renormalization of the anomalous average}
\label{App:Renormalize}
The generalized Gross-Pitaevskii equation [Eq.\ (\ref{Eq:GeneralizedGrossPitaevskiiEnd})] contains the anomalous average $\langle \tilde{\Lambda}(\mathbf{r})\tilde{\Lambda}(\mathbf{r})\rangle$, which is ultra-violet divergent.  We give a brief summary of the reason and cure for this problem  \cite{Hutchinson2000,Morgan2000,Rusch1999,Stoof1993,Proukakis1998,Kohler2002,Bijlsma1997}.

The divergence arises from of the use of the contact potential approximation. A genuinely ab initio theory would start by describing particle interactions using the true two-body potential. The contact ``potential'' is rather the zero-momentum limit of the two-body T-matrix describing the scattering of two particles in vacuum. It is introduced at  the outset [Eq.\ (\ref{Eq:FieldHamiltonian})] for a number  of reasons: partly for convenience, partly because this is the experimentally relevant quantity, and partly because it makes sense to include as much two-body physics as possible before embarking on a difficult many-body calculation. We certainly cannot treat the two-body interaction with perturbation theory. This is 
apparent from the fact that the interactions can be described by a contact potential dependent only on a scattering length, whereas a perturbative treament would depend on the details of the potential \cite{NoteThis}. 

However, this does mean that we have implicitly included at the outset various physical effects which must also appear in the many-body treatment. To avoid double-counting we need to subtract off the perturbative approximation to the two-body effects whenever we encounter them.  

The leading order interaction term is the nonlinear term involving the condensate. The interaction strength $U_0$ in this expression must now be replaced by the second order approximation, i.e.,
the $\tilde{U}$ in $\tilde{U}|\phi(\mathbf{r})|^{2}\phi(\mathbf{r})$ must be replaced in Eq.\ (\ref{Eq:GeneralizedGrossPitaevskiiEnd}) by
$\tilde{U}+\Delta \tilde{U}/N_{c}$, 
where
\begin{equation}
 \Delta \tilde{U} = \frac{\tilde{U}^{2}}{(2\pi)^3} \int 
d^3 \mathbf{k} \frac{m}{(\hbar k)^2},
\end{equation}
and $\Delta \tilde{U}/N_{c}^{2}$ is the second order correction to the interaction 
strength as calculated from the Lippmann-Schwinger equation. This correction can be grouped with the term in the generalized Gross-Pitaevskii equation [Eq.\ (\ref{Eq:GeneralizedGrossPitaevskiiEnd})] involving the anomalous average. 
This leads to a finite renormalized anomalous average $\tilde{m}^{R}(\mathbf{r})$, defined by
\begin{equation}
\tilde{m}^{R}(\mathbf{r}) = \langle \tilde{\Lambda}(\mathbf{r})\tilde{\Lambda}(\mathbf{r})\rangle + \frac{\Delta \tilde{U}}{\tilde{U}} \phi(\mathbf{r})^{2}.
\end{equation}
It should therefore be implicitly assumed that the anomalous average $\langle \tilde{\Lambda}(\mathbf{r})\tilde{\Lambda}(\mathbf{r})\rangle$ appearing in Eq.\ (\ref{Eq:GeneralizedGrossPitaevskiiEnd}) is replaced by $\tilde{m}^{R}(\mathbf{r})$ to produce a consistent, renormalized generalized Gross-Pitaevskii equation.

\section{Spectral decomposition of $\mathcal{L}(\mathbf{r},\mathbf{r'})$}
\label{App:Spectral}

The treatment described in this appendix echoes that of Castin and Dum \cite{Castin1998}, and is included for the sake of completeness.  

We assume that $(u_{k}(\mathbf{r'}),v_{k}(\mathbf{r'}))$ is a right eigenstate of the operator $\mathcal{L}(\mathbf{r},\mathbf{r'})$ defined in Eq.\ (\ref{Eq:TDBDGSpinor2LCal}), with eigenvalue $\epsilon_{k}$.  This is equivalently stated by
\begin{equation}
\int d\mathbf{r'}\mathcal{L}(\mathbf{r},\mathbf{r'}) 
\left(
\begin{array}{c}
u_{k}(\mathbf{r'})
\\
v_{k}(\mathbf{r'})
\end{array}
\right)
=
\epsilon_{k}\left(
\begin{array}{c}
u_{k}(\mathbf{r})
\\
v_{k}(\mathbf{r})
\end{array}
\right).
\label{Eq:LCalRightEig}
\end{equation}
Decomposing this two-component equation into the top and bottom elements then reveals, directly,
\begin{align}
\int d\mathbf{r'}L(\mathbf{r},\mathbf{r'})u_{k}(\mathbf{r'})
+\int d\mathbf{r'}M(\mathbf{r},\mathbf{r'})v_{k}(\mathbf{r'}) &= 
\epsilon_{k} u_{k}(\mathbf{r}),
\label{Eq:LRightEig}
\\
\int d\mathbf{r'}M^{*}(\mathbf{r},\mathbf{r'})u_{k}(\mathbf{r'}) 
+\int d\mathbf{r'}L^{*}(\mathbf{r},\mathbf{r'})v_{k}(\mathbf{r'}) &= 
-\epsilon_{k}v_{k}(\mathbf{r}).
\label{Eq:MStarRightEig}
\end{align}
Taking the complex conjugates of the above equations then yields
\begin{align}
\int d\mathbf{r'}L^{*}(\mathbf{r},\mathbf{r'})u_{k}^{*}(\mathbf{r'})
+\int d\mathbf{r'}M^{*}(\mathbf{r},\mathbf{r'})v_{k}^{*}(\mathbf{r'}) &= 
\epsilon_{k}^{*} u_{k}^{*}(\mathbf{r}),
\label{Eq:LStarRightEig}
\\
\int d\mathbf{r'}M(\mathbf{r},\mathbf{r'})u_{k}^{*}(\mathbf{r'}) 
+\int d\mathbf{r'}L(\mathbf{r},\mathbf{r'})v_{k}^{*}(\mathbf{r'}) &=  
-\epsilon_{k}^{*} v_{k}^{*}(\mathbf{r}).
\label{Eq:MRightEig}
\end{align}

Combining Eq.\ (\ref{Eq:LStarRightEig}) and Eq.\ (\ref{Eq:MStarRightEig}) then yields a ``left-hand'' form of Eq.\ (\ref{Eq:LCalRightEig}):
\begin{equation}
\int d\mathbf{r}
(u_{k}^{*}(\mathbf{r}), -v_{k}^{*}(\mathbf{r}))
\mathcal{L}(\mathbf{r},\mathbf{r'}) 
=
\epsilon_{k}^{*}
(u_{k}^{*}(\mathbf{r'}), -v_{k}^{*}(\mathbf{r'})).
\label{Eq:LCalLeftEig}
\end{equation}
We now choose a normalization convention for the two-component eigenstates such that
\begin{equation}
\int d\mathbf{r}
[|u_{k}(\mathbf{r})|^{2}
-
|v_{k}(\mathbf{r})|^{2}]
=1.
\label{Eq:Normalization}
\end{equation}
Hence, applying Eq.\ (\ref{Eq:LCalLeftEig}) onto a right eigenstate, where we can choose whether $\mathcal{L}(\mathbf{r},\mathbf{r'})$ should act to the right [Eq.\ (\ref{Eq:LCalRightEig})] or the left [Eq.\ (\ref{Eq:LCalLeftEig})], reveals that
\begin{equation}
\iint d\mathbf{r}d\mathbf{r'}
(u_{k}^{*}(\mathbf{r}), -v_{k}^{*}(\mathbf{r}))
\mathcal{L}(\mathbf{r},\mathbf{r'}) 
\left(
\begin{array}{c}
u_{k}(\mathbf{r'})
\\
v_{k}(\mathbf{r'})
\end{array}
\right)
=
\epsilon_{k}
=
\epsilon_{k}^{*},
\label{Eq:InnerProduct}
\end{equation}
i.e., the eigenvalue $\epsilon_{k}$ is real.

Thus, $(u_{k}^{*}(\mathbf{r}), -v_{k}^{*}(\mathbf{r}))$ is the corresponding left eigenstate, with eigenvalue $\epsilon_{k}$, to the right eigenstate appearing in Eq.\ (\ref{Eq:LCalRightEig}).  

Furthermore, Eq.\ (\ref{Eq:LStarRightEig}) and Eq.\ (\ref{Eq:MStarRightEig}) imply that the complex conjugate of a right eigenstate is also a right eigenstate:
\begin{equation}
\int d\mathbf{r'}\mathcal{L}(\mathbf{r},\mathbf{r'}) 
\left(
\begin{array}{c}
v_{k}^{*}(\mathbf{r'})
\\
u_{k}^{*}(\mathbf{r'})
\end{array}
\right)
=
-\epsilon_{k}\left(
\begin{array}{c}
v_{k}^{*}(\mathbf{r})
\\
u_{k}^{*}(\mathbf{r})
\end{array}
\right)
\label{Eq:LCalRightEigStar}
\end{equation}
Now, in an equivalently manner to the derivation of Eq.\ (\ref{Eq:LCalLeftEig}), from Eq.\ (\ref{Eq:LRightEig}) and Eq.\ (\ref{Eq:MRightEig}) we deduce that
\begin{equation}
\int d\mathbf{r}
(-v_{k}(\mathbf{r}), u_{k}(\mathbf{r}))
\mathcal{L}(\mathbf{r},\mathbf{r'}) 
=
-\epsilon_{k}
(-v_{k}(\mathbf{r'}), u_{k}(\mathbf{r'})),
\label{Eq:LCalLeftEigStar}
\end{equation}
i.e., that $(-v_{k}(\mathbf{r}), u_{k}(\mathbf{r}))$ is the corresponding left eigenstate, with eigenvalue $-\epsilon_{k}$, to the right eigenstate appearing in Eq.\ (\ref{Eq:LCalRightEigStar}).  

As the eigenstates have different eigenvalues, they are orthogonal, i.e.,
\begin{align}
\int d\mathbf{r}
[u_{k'}^{*}(\mathbf{r})u_{k}(\mathbf{r})
-
v_{k'}^{*}(\mathbf{r})v_{k}(\mathbf{r})]
= & \delta_{kk'},\\
\int d\mathbf{r}
[u_{k'}(\mathbf{r})v_{k}(\mathbf{r})
-
v_{k'}(\mathbf{r})u_{k}(\mathbf{r})]
=& 0
\end{align}
We note that setting $u_{k}(\mathbf{r})=\phi(\mathbf{r})$ and $v_{k}(\mathbf{r})=0$ on the one hand, and $v_{k}^{*}(\mathbf{r})=0$  and $u_{k}^{*}(\mathbf{r})=\phi^{*}(\mathbf{r})$ on the other, produces two eigenstates of eigenvalue zero. 

The identity can thus be decomposed as
\begin{equation}
\begin{split}
\delta(\mathbf{r}-\mathbf{r'})
\left(
\begin{array}{cc}
1 & 0
\\
0 & 1
\end{array}
\right)
=&
\left(
\begin{array}{c}
\phi(\mathbf{r})
\\
0
\end{array}
\right)
(
\phi^{*}(\mathbf{r'}), 0
)
+
\left(
\begin{array}{c}
0
\\
\phi^{*}(\mathbf{r})
\end{array}
\right)
(
0, \phi(\mathbf{r'})
)
\\&
+
\sum_{k=1}^{\infty}
\left(
\begin{array}{c}
u_{k}(\mathbf{r})
\\
v_{k}(\mathbf{r})
\end{array}
\right)
(
u_{k}^{*}(\mathbf{r'}), -v_{k}^{*}(\mathbf{r'})
)
\\&+
\sum_{k=1}^{\infty}
\left(
\begin{array}{c}
v_{k}^{*}(\mathbf{r})
\\
u_{k}^{*}(\mathbf{r})
\end{array}
\right)
(
-v_{k}(\mathbf{r'}), u_{k}(\mathbf{r'})
),
\end{split}
\end{equation}
and, similarly, $\mathcal{L}(\mathbf{r},\mathbf{r'})$ can be expressed as
\begin{equation}
\begin{split}
\mathcal{L}(\mathbf{r},\mathbf{r'})=&
\sum_{k=1}^{\infty}
\epsilon_{k}
\left(
\begin{array}{c}
u_{k}(\mathbf{r})
\\
v_{k}(\mathbf{r})
\end{array}
\right)
(
u_{k}^{*}(\mathbf{r'}), -v_{k}^{*}(\mathbf{r'})
)
\\&
-
\sum_{k=1}^{\infty}
\epsilon_{k}
\left(
\begin{array}{c}
v_{k}^{*}(\mathbf{r})
\\
u_{k}^{*}(\mathbf{r})
\end{array}
\right)
(
-v_{k}(\mathbf{r'}), u_{k}(\mathbf{r'})
).
\end{split}
\end{equation}

This is the usual form of the spectral decompositions of the operator $\mathcal{L}(\mathbf{r},\mathbf{r'})$ \cite{Castin1998}.  The two-component modes involving the condensate mode [which are also eigenstates of  $\mathcal{L}(\mathbf{r},\mathbf{r'})$] do not explicitly appear as they have eigenvalue zero.

\end{appendix}


\begin{thebibliography}{99}
\bibitem{NoteTrivially}
Trivially, the zeroth-order dynamics predicted by the first-order Hamiltonian described in Section \ref{SubSec:FirstOrderHamiltonian}) can also be considered consistent in this sense, as there is no evolution of the fluctuation operators $\tilde{\Lambda}(\mathbf{r})$, and hence no change in $N_{c}$.
\end{thebibliography}
\end{document}